\documentclass[a4paper]{article}

\usepackage[margin=3.0cm]{geometry}
\usepackage{amsmath}
\usepackage{amssymb}
\usepackage{xfrac}
\usepackage{algorithm}
\usepackage{algpseudocode}
\usepackage{graphics}
\usepackage{subfig}
\usepackage{color}
\usepackage{comment}
\usepackage{epic}
\usepackage{fancyvrb}
\usepackage{enumitem}

\setlist[description]{leftmargin=\parindent,labelindent=\parindent}

\newif\ifcommentandrea
\commentandreatrue
\usepackage{calc}
\newcommand{\andrea}[1]{\ifcommentandrea{\textcolor{blue}{(andrea: #1)}}\else \fi}

\newif\ifcommentdimitris
\commentdimitristrue
\newcommand{\dimitris}[1]{\ifcommentdimitris{\textcolor{red}{(dimitris: #1)}}\else \fi}

\usepackage{soulutf8}

\newif\iftodo
\todotrue
\usepackage{calc}

\author{Gianluca Frison, Dimitris Kouzoupis, Tommaso Sartor, Andrea Zanelli, Moritz Diehl \\ \it \small University of Freiburg, Department of Microsystems Engineering (IMTEK) and Department of Mathematics \\ \rm \small email: name.surname@imtek.uni-freiburg.de}
\title{BLASFEO: basic linear algebra subroutines \\ for embedded optimization}

\begin{document}

\makeatletter
\renewcommand*\env@matrix[1][*\c@MaxMatrixCols c]{%
  \hskip -\arraycolsep
  \let\@ifnextchar\new@ifnextchar
  \array{#1}}
\makeatother

\maketitle

\thanks{\footnotesize This research was supported by the EU via ERC-HIGHWIND (259 166), ITN-TEMPO (607 957), ITN-AWESCO (642 682), by Ministerium f\"ur Wissenschaft, Forschung und Kunst Baden-Wuerttemberg (Az: 22-7533.-30-20/9/3), by Bundesministerium f\"ur Wirtschaft und Energie (eco4wind, 0324125B), by the DFG in context of the Research Unit FOR 2401, and by Det Frie Forskningsr\r{a}d (DFF - 6111-00398).}

\begin{abstract}

BLASFEO is a dense linear algebra library providing high-performance implementations of BLAS- and LAPACK-like routines for use in embedded optimization.
A key difference with respect to existing high-performance implementations of BLAS is that the computational performance is optimized for small to medium scale matrices, i.e., for sizes up to a few hundred.
BLASFEO comes with three different implementations: a high-performance implementation aiming at providing the highest performance for matrices fitting in cache, a reference implementation providing portability and embeddability and optimized for very small matrices, and a wrapper to standard BLAS and LAPACK providing high-performance on large matrices. 
The three implementations of BLASFEO together provide high-performance dense linear algebra routines for matrices ranging from very small to large. 
Compared to both open-source and proprietary highly-tuned BLAS libraries, for matrices of size up to about one hundred the high-performance implementation of BLASFEO is about 20-30\% faster than the corresponding level 3 BLAS routines and 2-3 times faster than the corresponding LAPACK routines.

\end{abstract}


\section{Introduction} \label{sec:intro}

This paper introduces BLASFEO (Basic Linear Algebra Subroutines For Embedded Optimization), a dense linear algebra (DLA) library aiming at providing high-performance implementations of BLAS- and LAPACK-like routines for use in embedded optimization. 
BLASFEO is an open-source software~\cite{BLASFEO2016}, released under the LGPL license.

The first part of the name, Basic Linear Algebra Subroutines, echoes BLAS, which stands for Basic Linear Algebra Subprograms~\cite{Lawson1979}. 
The new word "Subroutines" indicates a key implementation feature of BLASFEO: the use of a modular design, based on assembly subroutines (as explained in detail in Section~\ref{sec:impl_det:subrou}). 
The second part of the name, For Embedded Optimization, is chosen to indicate the intended application area of the software. 
The assumptions behind the term embedded optimization are explained in detail in Section~\ref{sec:embopt}, and they affect many specific implementation choices, such as the focus on small-scale performance and the use of a non-conventional matrix format (referred to as panel-major in Section~\ref{sec:impl_det:panel}). 
This matrix format resembles the packed format of the internal memory buffers used in many BLAS implementations~\cite{Goto2008,Zee2015}.

The acronym, BLASFEO, reminds one of the word blasphemous, and in that it jokes about 
the choice of not using the canonical BLAS and LAPACK application programming interface (API) based on the column-major matrix format.
This choice is necessary in order to avoid the on-line conversion between the standard column-major matrix format and the panel-major matrix format, whose quadratic cost can not be well amortized in the case of small matrices.
For this reason, BLASFEO is not another BLAS implementation.

The primary aim of BLASFEO is to provide a DLA library to close the performance gap left by optimized BLAS and LAPACK implementations in the case of relatively small matrices, of size up to a few hundred.
The primary design goal of optimized BLAS and LAPACK implementations is to maximize throughput for large matrices.
This often comes at the cost of neglecting or even sacrificing small-scale performance.
To the best of our knowledge, there is no existing DLA library that aims at enhancing as much as possible the performance of DLA routines for relatively small matrices. 
Alternative approaches for small-scale DLA are in the direction of code generation \cite{Houska2011,Heinecke2016}, C++ templates \cite{eigen-hp} or specialized compilers \cite{Spampinato2016}.

BLASFEO comes with three implementations, introduced in Section~\ref{sec:impl}: a high-performance implementation (BLASFEO HP, aiming at providing the highest performance for matrices fitting in cache and employing hand-crafted assembly-coded DLA kernels), a reference implementation (BLASFEO RF, with portability and embeddability as main design goals, entirely coded in ANSI~C and optimized for very small matrices), and a wrapper to standard BLAS and LAPACK (BLASFEO WR, which ensures that BLASFEO performs no worse than optimized BLAS and LAPACK libraries and allows its performance to scale to large matrices).
{\color{black}
The BLASFEO HP and RF versions currently provide only single-threaded routines; the BLASFEO WR version can be linked to multi-threaded BLAS and LAPACK implementations.
}
In order to provide a unified framework that encompasses both the panel-major matrix format used in BLASFEO HP as well as the column-major format used in standard BLAS and LAPACK libraries, BLASFEO abstracts matrix and vector types by means of C structures (Section~\ref{sec:inter:str}).
Hence the use of a different API than BLAS and LAPACK (Section~\ref{sec:inter:api}).

\begin{figure}
\centering
\subfloat[{\tt dgemm\_nt}]{\includegraphics[width=0.42\linewidth]{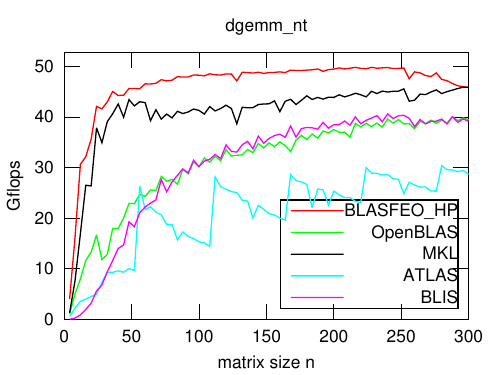} \label{fig:intro:dgemm_nt}} 
\subfloat[{\tt dpotrf\_l}]{\includegraphics[width=0.42\linewidth]{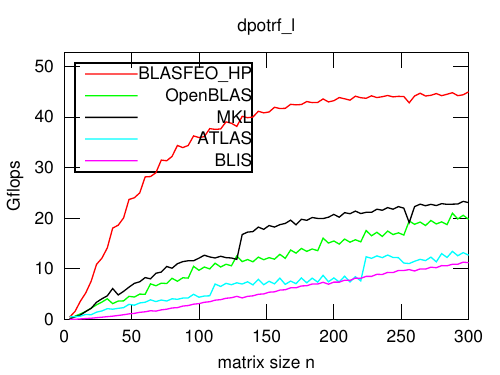} \label{fig:intro:dpotrf}} 
\caption{Performance of BLASFEO HP, OpenBLAS 0.2.19, MKL 2017.2.174, ATLAS 3.10.3 and BLIS 0.1.6 on one core of an Intel Core i7 4800MQ (Haswell architecture).}
\label{fig:intro}
\end{figure}

The main contribution of BLASFEO to the state-of-the-art in DLA is the level of performance reached by the BLASFEO HP implementation with respect to the corresponding BLAS (Figure~\ref{fig:intro:dgemm_nt}) and especially LAPACK routines (Figure~\ref{fig:intro:dpotrf}), in the case of matrix sizes up to a few hundred.
This performance level is due 
{\color{black}
to the fact that the panel-major matrix format is exposed to the user, as well as
}
to the careful choice and balance between many implementation techniques commonly employed in high-performance DLA routines, tailored to enhance small-scale performance.

At its core, BLASFEO HP employs hand-crafted assembly-coded DLA kernels with an API similar to the micro-kernel of BLIS~\cite{Zee2015}.
These kernels consist of the single innermost loop and employ register blocking (Section~\ref{sec:impl_det:block}) and vector instructions (Section~\ref{sec:impl_det:vec}).
However, BLIS implements a single micro-kernel (the nontransposed-transposed version of the matrix-matrix multiplication {\tt gemm\_nt}), and employs portable C-coded packing routines to handle the edge cases of other level 3 BLAS by means of properly padding and copying/transposing matrices while packing them into the internal optimized format.
This approach, which reduces the amount of assembly code and the object code size, has proven to be effective for large matrices, but it gives particularly poor performance for small ones (as it can be seen in Figure~\ref{fig:intro:dgemm_nt}, the performance ramps up much more slowly for matrices up to about 50), especially in the case of LAPACK routines (Figure~\ref{fig:intro:dpotrf}).

BLASFEO HP does not pack any data on-line (meaning every time a DLA routine is called; see Section~\ref{sec:embopt} for a generic definition of on-line and off-line in the context of embedded optimization), since it makes use of the panel-major matrix format (Section~\ref{sec:impl_det:panel}) that gives nearly optimal performance for matrices fitting in cache.
{\color{black}
Note that MKL supports offline packing for {\tt gemm} (with the routines {\tt gemm\_alloc}, {\tt gemm\_pack}, {\tt gemm\_compute} and {\tt gemm\_free}).
However, this functionality is provided only for {\tt gemm} and the data layout of the packed matrices is opaque, so it can not be employed as the native matrix format for optimization algorithms.
}
Furthermore, BLASFEO HP does not employ cache blocking, but the use of the panel-major matrix format together with a proper ordering of the two loops around the micro-kernel (Section~\ref{sec:impl_det:out_loop}) gives nearly optimal performance for matrices of size up to a few hundred (e.g., in Figure~\ref{fig:intro:dgemm_nt}, the performance is steadily close to the peak for matrix sizes up to 256).
Edge cases are handled explicitly by specialized DLA kernels.
Assembly code bloat is avoided by exploiting the modularity that can be achieved using assembly subroutines (Section~\ref{sec:impl_det:subrou}).
A tailored calling convention allows to split DLA kernels into small subroutines that perform elementary operations like loading/storing a sub-matrix to/from registers, or factorizing a register-resident sub-matrix.
DLA kernels are simply coded by combining subroutines like building blocks, 
and taking care of the specific function calling convention of the operating system (OS).
In this framework, the BLIS micro-kernel itself is reduced to an assembly subroutine, which is called by both level 3 BLAS and LAPACK DLA kernels (Section~\ref{sec:impl_det:lapack}).
Therefore LAPACK-like routines are not built on top of BLAS-like ones, but as if they were BLAS-like routines themselves.
This is a key difference with respect to the standard LAPACK implementation, greatly enhancing small-scale performance, as clearly visible in Figure~\ref{fig:intro:dpotrf}. 
Finally, the use of custom DLA routines can further enhance performance in the implementation of some algorithms (Section~\ref{sec:impl_det:custom}). 

The implementation approach employed in BLASFEO HP builds on the experience gained in the development of the DLA routines in HPMPC~\cite{Frison2014}, an interior point method for model predictive control described in the PhD thesis~\cite{Frison2015a}. 
The development of BLASFEO was motivated by the wish to make the DLA performance gains observed in HPMPC~\cite{Vukov2015} accessible to other software, in particular in the field of embedded optimization.

\section{Framework%
: embedded optimization} \label{sec:embopt}

Embedded optimization can be defined as the use of numerical optimization algorithms on embedded platforms for optimal decision making.
In embedded optimization, the distinction between on-line and off-line computations is important, the former being performed in-the-loop at every sampling time as soon as a new system measurement is available, the latter comprising pre-processing steps that can be precomputed ahead of time.
In embedded optimization, the optimization problems must be solved on-line, typically at high sampling frequencies in real-time, often on resource-constrained hardware.
A typical example is model predictive control (MPC)~\cite{Maciejowski2002}, a model-based advanced control technique that requires the solution of structured, constrained optimization problems at sampling times as low as in the microsecond range~\cite{Quirynen2015c,Jerez2014}.
This poses interesting challenges on the development of fast solvers for embedded optimization~\cite{Ferreau2017}.

Linear algebra routines are a key aspect in the implementation of these solvers, since they perform the most computationally expensive operations.
This paper focuses on level 3 BLAS- and LAPACK-like routines, which are the backbone of second-order optimization methods, i.e. algorithms that make use up to second-order derivative information to solve the optimization problem.
Level 2 BLAS-like routines, which are the backbone of first order optimization methods, are beyond the scope of this paper.
A set of linear algebra routines tailored to embedded optimization problems can take advantage of the special features of this class of problems in order to reduce the computational time.
The following features are considered:
\begin{enumerate}
\item Embedded optimization problems must be solved in real-time, often on resource-constrained hardware.
The computational speed is a key factor.
\item The size of the matrices is often relatively small, i.e., in the order of tens up to a few hundred.
Embedded optimization problems can have several thousands of variables, but they are often structured and therefore the optimization algorithms can exploit this structure and perform computations on smaller matrices.
\item Structure-exploiting optimization algorithms can exploit the high-level sparsity pattern of the problem and therefore the data matrices are generally dense. 
\item Numerical optimization algorithms are typically iterative schemes.  
Furthermore, a sequence of similar problems is solved at each sampling time.
This implies that each data matrix is typically reused several times. 
\end{enumerate}

These features can be exploited in the design of linear algebra routines as follows:
\begin{enumerate}

\item Linear algebra routines must make an efficient use of available hardware resources.
Compilers often do a poor job in converting generic triple-loop based linear algebra source code into efficient object code fully exploiting hardware capabilities, especially if the hardware lacks features like out-of-order execution and hardware prefetch.
Therefore high-performance implementation techniques should be employed in the implementation of fast linear algebra routines. 


\item Matrices with sizes in the order of tens or a few hundred are assumed to fit in some cache level.
As a consequence, implementation techniques like cache blocking are not considered, simplifying the design of the linear algebra routines.
Furthermore, for small matrices the cost of packing is not negligible with respect to the cost of performing level 3 BLAS operations.
Therefore, linear algebra routines should be designed to reduce as much as possible the need of copying data.


\item Sparse linear algebra requires the use of special matrix formats and the efficient handling of matrix element indices.
Sparse linear algebra can make limited use of processor features like vectorization, and therefore it has lower computational throughput, limiting its use only to the case of very sparse problems.
Therefore, only dense linear algebra routines are considered, with the exception of very special and common sparse matrices with fixed structure (i.e., diagonal or triangular).

\item Since data matrices are typically reused several times, it makes sense to store them in a matrix format that is particularly favorable for the linear algebra routines.
The cost to convert matrices into this format can be amortized over several matrix reuses, or the conversion may even be performed off-line for data that does not depend on the system measurements.


\end{enumerate}

\section{Implementations} \label{sec:impl}

This section briefly describes the three BLASFEO implementations.
Section~\ref{sec:impl_det} contains more details about the techniques used in order to obtain high-performance routines in the BLASFEO HP and RF implementations.

{\color{black}
All BLASFEO implementations share a common API.
This API is based on C structures for matrices and vectors, and therefore it differs from the API of BLAS and LAPACK.
This more object-oriented approach is necessary to conveniently hide implementation details, as BLASFEO deals with different matrix formats (column-major for BLASFEO RF and BLASFEO WR, and panel-major for BLASFEO HP).

In all BLASFEO routines, each matrix is described by means of the three arguments
\begin{verbatim}
struct blasfeo_dmat *sA, int ai, int aj
\end{verbatim}
where {\tt sA} is a pointer to a (double-precision) matrix structure, and {\tt ai} and {\tt aj} are the row and column index of the top left corner of the working sub-matrix, respectively.
Each BLASFEO implementation comes with its own implementation of the matrix structure, whose internal details are hidden to the user.
More details about the BLASFEO API can be found in Section~\ref{sec:inter} of the appendix.
}

\subsection{BLASFEO WR} \label{sec:impl:rf}

The wrapper version of BLASFEO (BLASFEO WR) provides a thin wrapper to the Fortran version of BLAS and LAPACK routines.
It allows one to automatically port BLASFEO to each new architecture for which a BLAS version exists.
Furthermore, by linking to optimized BLAS implementations it gives good performance for large matrices, and the possibility to exploit \mbox{multi-core} CPUs.

{\color{black}
In BLASFEO WR, the {\tt blasfeo\_dmat} structure has a member {\tt pA} (of type pointer to {\tt double}) pointing to the first element of a matrix in column-major matrix format.
The member {\tt m} (of type {\tt int}) provides the matrix leading dimension in the BLAS notation.
}
The wrapper simply takes care of extracting this information and updating the pointer to the first element of the working sub-matrix, as
\begin{verbatim}
int lda   = sA->m;
double *A = sA->pA + ai + aj*lda;
\end{verbatim}
where {\tt sA} is a pointer to a {\tt blasfeo\_dmat} structure, and {\tt ai} and {\tt aj} are the coordinates of the first element of the sub-matrix that is the actual operand.

Optionally, the wrapper can perform additional consistency checks before calling the BLAS or LAPACK routine, for example to make sure that the operation is not exceeding the boundaries of the matrix.

\subsection{BLASFEO RF} \label{sec:impl:rf}

The reference implementation of BLASFEO (BLASFEO RF) has the aim of providing a rather concise and machine-independent implementation, performing well for very small matrices.

Like BLASFEO WR, it makes use of the column-major matrix format.
Therefore, the first element of each working sub-matrix is again computed as
\begin{verbatim}
int lda   = sA->m;
double *A = sA->pA + ai + aj*lda;
\end{verbatim}
%
BLASFEO RF is written in ANSI C code, without any use of machine-specific instructions or intrinsics.
The code is slightly optimized, with high performance for very small matrices and the widest machine compatibility in mind.
In the code optimization, it is assumed that the target machine has at least 8 scalar floating-point (FP) registers.

Each level 3 BLAS routine is written as 3 nested loops.
The innermost loop is over $k$, and therefore it performs dot products.
Cache blocking is not employed.
Register blocking is employed (Section \ref{sec:impl_det:block}), with $2\times 2$ block size.
Therefore, it is assumed that 4 FP registers are used to hold the $2\times 2$ sub-matrix of the result matrix, while the remaining FP registers are used to hold elements from the factor matrices and intermediate results.
The use of $2\times 2$ register block size provides a reuse factor of 2 of elements from the factor matrices.
Furthermore, it provides 4 independent accumulators, helping hiding the latency of FP operations.

\subsection{BLASFEO HP} \label{sec:impl:hp}

The high-performance implementation of BLASFEO (BLASFEO HP) has the aim of providing linear algebra routines with the highest computational performance, assuming that matrices fit in some cache level. 


BLASFEO HP does not make use of cache blocking.
Therefore, level 3 linear algebra routines are implemented using three nested loops.
The inner most loop is coded in C or assembly, hand-optimized for the target architecture and operating system (Section \ref{sec:impl_det:target}). 
Register blocking is employed (Section \ref{sec:impl_det:block}), with blocking size depending on the target architecture.
Vectorization is employed thanks to the explicit use of SIMD instructions (Section \ref{sec:impl_det:vec}), again depending on the target architecture.
Matrices are stored in panel-major format (Section \ref{sec:impl_det:panel}).
This format is analogous to the packed matrix format internally used in GotoBLAS/OpenBLAS/BLIS.
{\color{black}
The order of the two outer loops has to be chosen properly (Section \ref{sec:impl_det:out_loop}), as it affects the cache bandwidth requirements.
}
Linear algebra kernels are coded in assembly in a modular fashion, making heavy use of subroutines with custom calling convention (Section \ref{sec:impl_det:subrou}): corner cases are implemented as a trade-off between code size and performance. 
There exist specialized kernels for each linear algebra routine, and in particular LAPACK routines are implemented as if they were level 3 BLAS routines, and not on top of them (Section \ref{sec:impl_det:lapack}).
As a further step in the same direction, the use of custom DLA routines can merge several routines into a single one, reducing overhead in the case of small matrices (Section \ref{sec:impl_det:custom}).

\section{Details of high-performance implementation techniques} \label{sec:impl_det}

This section presents the details of the high-performance techniques used in the implementation mainly of BLASFEO HP and, to a smaller extent, of BLASFEO RF.
Most techniques are standard practice in high-performance BLAS implementations, but they are revised in the embedded optimization framework.
{\color{black}
In particular, the choice to avoid on-line packing and expose the panel-major matrix format creates major implementation challenges.
BLASFEO HP is the result of careful implementation trade-offs, necessarily leading to a design that is slightly sub-optimal in some aspects, but with excellent performance in practice.
}

Sections~\ref{sec:impl_det:block} to \ref{sec:impl_det:subrou} describe the implementation of the {\tt gemm} kernel, which is the backbone of all level 3 BLAS and LAPACK routines.
Indeed, the computationally most expensive parts of all level 3 BLAS routines can be cast in terms of this kernel~\cite{Kaagstroem1998,Goto2008a}.
In turn, in standard implementations LAPACK routines are built on top of level 3 BLAS routines, and therefore the {\tt gemm} kernel accounts for most of the computations in LAPACK routines, too.
Sections~\ref{sec:impl_det:lapack} and \ref{sec:impl_det:custom} apply the proposed implementation scheme to other level 3 BLAS and LAPACK routines with focus on small-scale performance.


\subsection{Register blocking} \label{sec:impl_det:block}

Register blocking is the simultaneous computation of all elements of a sub-matrix (or block) of the result matrix fitting into registers.
It has the twofold aim of hiding latency of instructions, and of reducing the number of memory operations. 

\subsubsection{Hiding instruction latency} 

In modern computer architectures, most FP instructions are pipelined.
The execution of a pipelined instruction is split into stages.
While an instruction is at a certain stage of the pipeline, other instructions can be processed at the same time, at other stages of the pipeline.
Therefore the instruction latency (defined as the number of clock cycles for the result of the instruction to be available as an input to other instructions) is larger than the instruction throughput (defined as the reciprocal of the maximum number of such instructions that can be processed per clock cycle).
If a code fragment containts a long sequence of equal and independent instructions, after an initial delay, equal to the instruction latency, all stages of the pipeline are busy working on different instructions, and an instruction is processed every number of clock cycles equal to the instruction throughput.
If there is dependency between the output of an instruction and the input of a following instruction, then the second instruction cannot be processed until the result of the first instruction is available: this stalls the pipeline.

Register blocking can be used to hide instruction latency.
{\color{black}
In particular, in order to code a high-performance {\tt gemm} kernel, the pipelines for FP multiplication and addition (or for fused-multiplication-accumulation, depending on the architecture) must be kept as busy as possible.
}
The computation of several matrix elements at the same time can provide enough independent instructions to keep 
{\color{black}
these pipelines
}
fully utilized.

\subsubsection{Reducing the number of memory operations} 

Register blocking allows one to reuse each matrix element several times once it is loaded into registers
{\color{black}
(i.e., increase arithmetic intensity).
}
Therefore, fewer memory operations are necessary to perform the same number of flops.
This is useful to reduce the memory bandwidth requirements below the maximum memory bandwidth available in the system, and therefore to avoid that the DLA kernels become memory-bounded.

The blocking idea can generally be applied to other memory levels (as for example cache blocking) to take into account the fact that the available memory bandwidth typically decreases at lower levels in the memory hierarchy. 
However, since BLASFEO HP and RF target relatively small matrices that are assumed to fit in 
{\color{black}
the last level of cache (LLC),
}
cache blocking is not employed in their implementation.
Therefore, their performance deteriorates for larger matrices.

{\color{black}
\subsubsection{Remark on arithmetic intensity}
Register blocking size affects arithmetic intensity.
As a minimum, this has to be large enough to allow the data streamed from L1 cache to keep the computational kernel fed (this typically means to not exceed the load instructions throughput).
BLASFEO HP targets matrices fitting in LLC and it does not employ cache blocking, therefore bandwidth with respect to main memory is not of interest.
Section~\ref{sec:impl_det:out_loop} proposes an order for the two outer loops (such that the L1 cache contains a sub-matrix of the left factor $A$, while the right factor $B$ is streamed from L2 or L3 cache) that minimizes the amount of data streamed from LLC, attempting to keep the bandwidth requirements below the LLC maximum bandwidth.
The numerical experiments in Section~\ref{sec:exp} show that the performance of BLASFEO HP is typically very close to the maximum throughput already for rather small matrices, hinting that on the target architectures the BLASFEO HP routines are compute bounded and not memory bounded for matrices fitting in LLC.
For very small matrices (fitting in L1 cache), the performance limiting factor is the computation overhead (due to the logic to select the correct DLA kernel sizes and the function calling overhead, as well as to the $\mathcal O(n^2)$ and $\mathcal O(n)$ terms in the DLA kernel, like e.g. scaling of the result matrix in {\tt gemm}, or using division instructions in factorizations), rather than the insufficient memory bandwidth.
}

\subsection{Vectorization} \label{sec:impl_det:vec}

Vectorization is the redesign of an algorithm to take advantage of the vector processing capability of the hardware.
Many modern architectures feature Single-Instruction Multiple-Data (SIMD) instructions that perform the same operation in parallel on all elements of small vectors of data. 
In theory, instructions operating on vectors of size $n_v$ can boost the performance up to a factor $n_v$.
SIMD are an easy and efficient way to increase single-thread performance, especially in scientific computing.

As an example, the x86 and x86\_64 architectures have several versions of SSE instructions (operating on 128-bit-wide vectors, each holding 2 double or 4 single precision FP numbers) and AVX instructions (operating on 256-bit-wide vectors, each holding 4 double or 8 single precision FP numbers), while the ARM architecture has NEON instructions (operating on 128-bit-wide vectors). 

Compilers can attempt to automatically vectorize scalar code, emitting SIMD instructions.
However, producing efficient SIMD code is not a simple task, since it may require deep changes to the code structure
that are often better suited to the programmer, who has a better high-level overview of the algorithm. 
{\color{black}
As an example, the {\tt gemm} kernel in optimized BLAS implementations has the inner loop over $k$ (corresponding to a {\tt dot} operation), and the two loops around it are partially unrolled in order to block for cache and allow vectorization.
However, compilers often seem unable to perform such code transformations, and perform better with an inner loop over $i$ (corresponding to an {\tt axpy} operation), which is partially unrolled by the compiler to provide vectorization.
See Section~\ref{sec:compilers} of the appendix for a performance comparison between some popular compilers in optimizing Netlib-style DLA.
}

The use of SIMD can be ensured by explicitly coding them in assembly or inline assembly (low level solution, giving full control also over the instruction scheduling and register allocation) or by means of intrinsics (higher lever solution, where intrinsics are special functions called from C code and directly mapped to SIMD instructions, leaving instruction scheduling and register allocation to the compiler).
BLASFEO RF does not make explicit use of vectorization, while BLASFEO HP uses assembly-coded DLA kernels in order to have access to the entire instruction set and have full control over register allocation and instruction scheduling.

\subsection{Panel-major matrix format} \label{sec:impl_det:panel}

{\color{black}
The use of contiguous memory is a key factor in the implementation of high-performance DLA routines~\cite{Henry1992}.
E.g. it helps
}
to fully exploit the available memory bandwidth, it improves cache reuse and it reduces the Translation Lookaside Buffer (TLB) misses. 

\subsubsection{Use of contiguous memory}

When an element is fetched from memory, data is moved into cache in chunks (called cache lines) of typically 32 or 64 bytes. 
This means that the access to elements belonging to the same cache line is faster, since only one cache line needs to be moved into cache.
In contrast to that, random access of elements typically requires a different cache line for each element.
Therefore the access of contiguous elements maximizes the effective memory bandwidth.

In order to speed up cache access and reduce its complexity and cost, a certain cache line (depending on its memory address) can be mapped to a limited number $n$ of locations in cache: this kind of cache is called $n$-way associative.
Due to associativity, it may happen that cache lines are evicted from cache even if this is not fully utilized.
As an example, if a matrix is stored in column-major order, for certain column lengths it can happen that contiguous elements on the same row are mapped into the same cache location, evicting each other.
This effectively acts as a reduction in cache size.
Use of contiguous memory can mitigate this, since consecutive cache lines are mapped in different cache locations.

Finally, memory is seen from a program as virtual memory, 
that is mapped into physical memory locations by means of a translation table in the MMU (Memory Management Unit), the page table.
The TLB is a cache for the page table, containing the physical address of the most recently used memory pages (each usually of size 4 KB).
If memory is accessed in a non-contiguous way, it may happen that TLB is not large enough to translate the entire content of cache, increasing the number of expensive TLB misses.

\subsubsection{GotoBLAS approach}

In~\cite{Goto2002}, a {\tt gemm} design based on reducing TLB misses is proposed.
In this approach, 
{\color{black}
a multilayered blocking approach is employed.
The working
}
sub-matrices from the $A$ and $B$ matrices are packed into memory buffers
{\color{black}
$\widetilde A$ and $\widetilde B$
}
before each call to the {\tt gemm} kernel.
These sub-matrices are carefully packed (and possibly transposed)
{\color{black}
into row-panel and column-panel matrix formats respectively.
This approach
}
is employed also in OpenBLAS~\cite{OpenBLAS2011} and clearly presented in the BLIS paper~\cite{Zee2016}.

Matrix elements are stored in the exact same order as accessed by the {\tt gemm} kernel, and taking into account cache and TLB sizes and associativities.
{\color{black}
In the buffers $\widetilde A$ and $\widetilde B$, matrix elements are stored into fixed-size panels
}
(which are sub-matrices with many more rows than columns, or the other way around)
{\color{black}
of contiguous data.
The smaller size of each panel (in short, panel size) depends on the size $m_r \times n_r$ of the {\tt gemm} kernel, and it is equal to $m_r$ for $\widetilde A$ and $n_r$ for $\widetilde B$.
In the common case of $m_r\not = n_r$, the panel sizes for $\widetilde A$ and $\widetilde B$ are different.
}
The result matrix $C$ is stored in column-major format.

This approach gives near full FP throughput for large matrices, but it suffers from a severe overhead for small matrices, since in this case the (quadratic) cost of packing data can not be well amortized over the (cubic) cost of performing FP operations. 

\subsubsection{Panel-major matrix format}

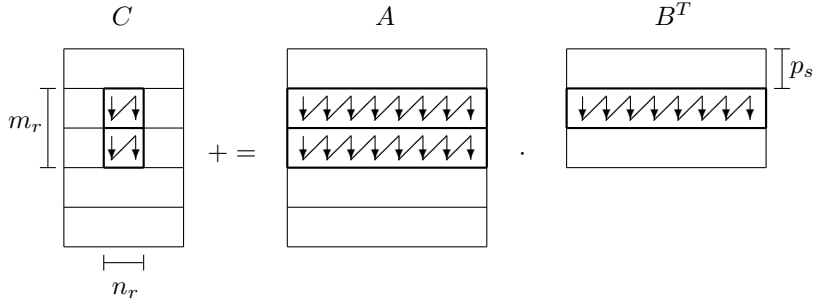
\begin{figure}
\centering
\setlength{\unitlength}{0.7\linewidth}
\begin{picture}(1, 0.4)

\thinlines
\put(0.07,0.35){\line(1,0){0.15}}
\put(0.07,0.30){\line(1,0){0.15}}
\put(0.07,0.25){\line(1,0){0.15}}
\put(0.07,0.20){\line(1,0){0.15}}
\put(0.07,0.15){\line(1,0){0.15}}
\put(0.07,0.10){\line(1,0){0.15}}
\put(0.07,0.10){\line(0,1){0.25}}
\put(0.22,0.10){\line(0,1){0.25}}
\thicklines
\put(0.12,0.30){\line(1,0){0.05}}
\put(0.12,0.25){\line(1,0){0.05}}
\put(0.12,0.20){\line(1,0){0.05}}
\put(0.12,0.20){\line(0,1){0.10}}
\put(0.17,0.20){\line(0,1){0.10}}
\thinlines
\put(0.13,0.29){\vector(0,-1){0.03}}
\drawline(0.13,0.26)(0.16,0.29)
\put(0.16,0.29){\vector(0,-1){0.03}}
\put(0.13,0.24){\vector(0,-1){0.03}}
\drawline(0.13,0.21)(0.16,0.24)
\put(0.16,0.24){\vector(0,-1){0.03}}

\put(0.25,0.21){$+=$}

\thinlines
\put(0.35,0.35){\line(1,0){0.25}}
\put(0.35,0.30){\line(1,0){0.25}}
\put(0.35,0.25){\line(1,0){0.25}}
\put(0.35,0.20){\line(1,0){0.25}}
\put(0.35,0.15){\line(1,0){0.25}}
\put(0.35,0.10){\line(1,0){0.25}}
\put(0.35,0.10){\line(0,1){0.25}}
\put(0.60,0.10){\line(0,1){0.25}}
\thicklines
\put(0.35,0.30){\line(1,0){0.25}}
\put(0.35,0.25){\line(1,0){0.25}}
\put(0.35,0.20){\line(1,0){0.25}}
\put(0.35,0.20){\line(0,1){0.10}}
\put(0.60,0.20){\line(0,1){0.10}}
\thinlines
\put(0.37,0.29){\vector(0,-1){0.03}}
\drawline(0.37,0.26)(0.40,0.29)
\put(0.40,0.29){\vector(0,-1){0.03}}
\drawline(0.40,0.26)(0.43,0.29)
\put(0.43,0.29){\vector(0,-1){0.03}}
\drawline(0.43,0.26)(0.46,0.29)
\put(0.46,0.29){\vector(0,-1){0.03}}
\drawline(0.46,0.26)(0.49,0.29)
\put(0.49,0.29){\vector(0,-1){0.03}}
\drawline(0.49,0.26)(0.52,0.29)
\put(0.52,0.29){\vector(0,-1){0.03}}
\drawline(0.52,0.26)(0.55,0.29)
\put(0.55,0.29){\vector(0,-1){0.03}}
\drawline(0.55,0.26)(0.58,0.29)
\put(0.58,0.29){\vector(0,-1){0.03}}
\thinlines
\put(0.37,0.24){\vector(0,-1){0.03}}
\drawline(0.37,0.21)(0.40,0.24)
\put(0.40,0.24){\vector(0,-1){0.03}}
\drawline(0.40,0.21)(0.43,0.24)
\put(0.43,0.24){\vector(0,-1){0.03}}
\drawline(0.43,0.21)(0.46,0.24)
\put(0.46,0.24){\vector(0,-1){0.03}}
\drawline(0.46,0.21)(0.49,0.24)
\put(0.49,0.24){\vector(0,-1){0.03}}
\drawline(0.49,0.21)(0.52,0.24)
\put(0.52,0.24){\vector(0,-1){0.03}}
\drawline(0.52,0.21)(0.55,0.24)
\put(0.55,0.24){\vector(0,-1){0.03}}
\drawline(0.55,0.21)(0.58,0.24)
\put(0.58,0.24){\vector(0,-1){0.03}}

\put(0.64,0.21){$\cdot$}

\thinlines
\put(0.70,0.35){\line(1,0){0.25}}
\put(0.70,0.30){\line(1,0){0.25}}
\put(0.70,0.25){\line(1,0){0.25}}
\put(0.70,0.20){\line(1,0){0.25}}
\put(0.70,0.20){\line(0,1){0.15}}
\put(0.95,0.20){\line(0,1){0.15}}
\thicklines
\put(0.70,0.30){\line(1,0){0.25}}
\put(0.70,0.25){\line(1,0){0.25}}
\put(0.70,0.25){\line(0,1){0.05}}
\put(0.95,0.25){\line(0,1){0.05}}
\thinlines
\put(0.72,0.29){\vector(0,-1){0.03}}
\drawline(0.72,0.26)(0.75,0.29)
\put(0.75,0.29){\vector(0,-1){0.03}}
\drawline(0.75,0.26)(0.78,0.29)
\put(0.78,0.29){\vector(0,-1){0.03}}
\drawline(0.78,0.26)(0.81,0.29)
\put(0.81,0.29){\vector(0,-1){0.03}}
\drawline(0.81,0.26)(0.84,0.29)
\put(0.84,0.29){\vector(0,-1){0.03}}
\drawline(0.84,0.26)(0.87,0.29)
\put(0.87,0.29){\vector(0,-1){0.03}}
\drawline(0.87,0.26)(0.90,0.29)
\put(0.90,0.29){\vector(0,-1){0.03}}
\drawline(0.90,0.26)(0.93,0.29)
\put(0.93,0.29){\vector(0,-1){0.03}}

\thinlines
\put(0.04,0.30){\line(1,0){0.02}}
\put(0.04,0.20){\line(1,0){0.02}}
\put(0.05,0.20){\line(0,1){0.10}}
\put(0.00,0.25){$m_r$}

\thinlines
\put(0.12,0.07){\line(0,1){0.02}}
\put(0.17,0.07){\line(0,1){0.02}}
\put(0.12,0.08){\line(1,0){0.05}}
\put(0.13,0.04){$n_r$}

\thinlines
\put(0.96,0.35){\line(1,0){0.02}}
\put(0.96,0.30){\line(1,0){0.02}}
\put(0.97,0.30){\line(0,1){0.05}}
\put(0.98,0.32){$p_s$}

\put(0.46,0.38){$A$}
\put(0.81,0.38){$B^T$}
\put(0.13,0.38){$C$}


\end{picture}

\caption{
Matrix layout in memory (called panel-major matrix format): matrix elements are stored in the same order as the {\tt gemm} kernel accesses them. 
This {\tt gemm} kernel implements the optimal 'NT' variant (namely, $A$ nontransposed, $B$ transposed), and it computes a $m_r \times n_r$ sub-matrix of the result matrix $C$.
The panel height $p_s$ is the same for the left and the right factors $A$ and $B$, as well as for the result matrix $C$.
Each arrow represents the $p_s$ elements that are on the same column within a panel.
The diagonal lines indicate that, in a contiguous memory swap, once the last element of a column is accessed, the following element to be accessed is the first element of the following column within the same panel.
}
\label{matrix_layout}
\end{figure}

{\color{black}
In embedded optimization, matrices are generally rather small, and assumed to fit in cache.
In this case, if only the three innermost loops in the GotoBLAS approach are considered and the matrices $A$ and $B$ are already in the form of the buffers $\widetilde A$ and $\widetilde B$, high-performance DLA routines can be obtained.
This approach requires to expose the row-panel and column-panel matrix formats used for $\widetilde A$ and $\widetilde B$, and to decouple the packing routines form the DLA routines.
In the context of embedded optimization, where optimization algorithms are typically iterative and therefore data matrices are reused several times, this approach has the advantage of allowing to well amortize packing cost also in case of small matrices.
Note however that in general $\widetilde A$ and $\widetilde B$ are stored into different formats (as e.g. the panel sizes $m_r$ and $n_r$ can differ), and that this severely limits the flexibility of the approach.
Furthermore, the output matrix is not in packed format, but in panel-major.

BLASFEO HP attempts to overcome these limitations by proposing a paneled matrix format (called panel-major) with fixed panel size for all input as well as output matrices.
}
%
More in details, in the {\tt gemm} routine, the $A$ and $B$ matrices are packed into horizontal panels of contiguous data, as shown in Figure~\ref{matrix_layout}.
{\color{black}
The panel size is fixed and denoted by $p_s$.
}
As a consequence, the DLA kernel size $m_r\times n_r$ is generally chosen such that both $m_r$ and $n_r$ are a multiple of $p_s$.
The values of $m_r$ and $n_r$ are architecture-dependent and a function of the number of registers as well as the SIMD width.
The value of $p_s$ is usually chosen as the smaller of $m_r$ and $n_r$, such that every time a cache line is accessed, it is fully utilized.
{\color{black}
The output matrix $C$ is directly stored in the panel-major format, avoiding the need for further packing.
The choice of fixing the panel size to the same value $p_s$ for all matrices greatly simplifies the implementation of DLA routines, and it allows each panel-major matrix to be freely used as an input or output of any DLA routine.
Therefore, the panel-major format is used as the native matrix format in BLASFEO HP.
}

{\color{black}
The panel-major matrix format can be seen as a three-dimensional array, where the size of one dimension is fixed to $p_s$.
In general DLA routines for tensors can be employed to operate on the panel-major matrix format.
However, for efficiency reasons, the fact that one dimension is fixed to $p_s$ is exploited in the implementation of the routines in BLASFEO HP.
}

In the panel-major matrix format, the first element of each working sub-matrix is computed differently than in the case of a column-major matrix.
The index of the panel containing the element is {\tt ai/ps}, which, multiplied by the panel length, gives the offset of the first panel element with respect to the first matrix element.
The column index of the element in the panel is {\tt aj}, which,  multiplied by the panel size {\tt ps}, gives the offset of the first column element with respect to the first panel element.
Finally, the row index of the element in the column is the reminder {\tt ai\%ps}.
In summary, the first element of the sub-matrix is
\begin{verbatim}
int sda   = sA->cn;
double *A = sA->pA + ai/ps*ps*sda + aj*ps + ai%ps;
\end{verbatim}
where {\tt sA} is a pointer to a {\tt blasfeo\_dmat} struct and {\tt ai} and {\tt aj} are the coordinates of the first element of the sub-matrix that is the actual operand.
The integer {\tt sda} (standing for second dimension of matrix A, analogous to {\tt lda} in standard BLAS, but referring to the other dimension) is the length of each panel, which can be larger than the number of columns {\tt n} if padding for alignment is employed.
The computation is efficiently implemented as
\begin{verbatim}
int sda   = sA->cn;
int air   = ai & (ps-1);
double *A = sA->pA + (ai-air)*sda + aj*ps + air;
\end{verbatim}
The operation
\begin{verbatim}
air = ai & (ps-1)
\end{verbatim}
is the reminder of the division of {\tt ai} by {\tt ps}, implemented exploiting the fact that {\tt ps} is a power of two, and therefore {\tt ps-1} can be used as a mask for the reminder. 
The meaning of each of the pointer updates is:
\begin{description}
\item[{\tt (ai-air)*sda}] is an efficient implementation of {\tt ai/ps*ps*sda}, where the operation {\tt ai/ps} computes the number of the panel where the ({\tt ai})-th row is; this is then multiplied by {\tt ps*sda}, the size (in doubles) of each panel.
\item[{\tt aj*ps}] is the position of the ({\tt aj})-th column in the ({\tt ai/ps})-th panel (that can be seen as a column-major matrix with {\tt lda} equal to {\tt ps}).
\item[{\tt air}] is the position of the ({\tt ai})-th row in the ({\tt ai/ps})-th panel.
\end{description}
It is important to note that the value of {\tt ps} is chosen as a power of 2 and it is defined as a constant: therefore the compiler knows its value and can e.g. implement multiplications as faster (arithmetic) shifts left.

Figure~\ref{matrix_layout} shows the panel-major matrix layout and the behavior of the 'NT' variant of the {\tt gemm} kernel that computes $D \gets \alpha \cdot A\cdot B^T + \beta \cdot C$, where the left factor $A$ is nontransposed and the right factor $B$ is transposed.
This is the optimal variant, since both $A$ and $B$ are accessed panel-wise (i.e. data is read along panels).
Furthermore, the regular access pattern of data in memory (i.e. access of contiguous memory locations) can be easily detected by the hardware prefetcher (if present in the architecture).

In the 'NN' variant of the {\tt gemm} kernel, the $A$ matrix is optimally accessed panel-wise, but the $B$ matrix is accessed across panels (i.e., only a few columns of each $B$ panel are used, before moving to the following panel) making a worse use of caches and TLBs.
This complex access pattern is generally not detected by the hardware prefetcher, and therefore software prefetch has to be explicitly used to move $B$ elements into cache before they are needed.

{\color{black}
The {\tt gemm} routine variants 'TN' and 'TT' (where the left matrix factor $A$ is transposed) are not implemented explicitly, due to the inefficient use of vector instructions in these schemes.
Both 'TN' and 'TT' variants of the {\tt gemm} operation can be implemented by explicitly transposing the matrix $A$ and therefore using the 'NN' and 'NT' variants of the {\tt gemm} routine.
Alternatively, a native 'TT' {\tt gemm} routine can be implemented by using the 'NN' assembly subroutine, and transposing the $m_r\times n_r$ result sub-matrix before storing it, as $A^T\cdot B^T = (B\cdot A)^T$.
}

\subsection{Assembly subroutines and modularity} \label{sec:impl_det:subrou}
In BLASFEO HP, the optimized linear algebra kernels are coded in assembly.
This choice has been made for several reasons.
%
%
One key reason is that assembly allows for much more modularity than what would be possible in higher level languages without compromising performance. 
Function calling conventions in high level languages severely limit the use of FP registers to pass data.
Therefore, it is not possible to split a linear algebra kernel into smaller functions without having to repeatedly store and load the data from accumulation registers, introducing a severe overhead.
Conversely, in assembly it is possible to split a linear algebra kernel into subroutines (that is, into blocks of code that perform specific tasks) and to have complete freedom in the definition of more convenient calling conventions.

\subsubsection{Subroutines and custom calling convention}

In BLASFEO HP each assembly module contains a number of subroutines with local scope performing basic operations (e.g., in the implementation of the Cholesky factorization: the {\tt gemm} loop, the loading of a sub-matrix, the Cholesky factorization of a register-resident matrix, the storing of the result).
Modules also contain functions with global scope (and therefore following the OS calling convention) that simply consist of the glue between a sequence of calls to the subroutines.
A custom calling convention for the subroutines allows to pass data in FP registers between different subroutines and avoids the overhead of standard calling conventions (e.g. store of registers on the stack in the prologue, load of registers from the stack in the epilogue).
%
Therefore, DLA kernels are built in a modular fashion by using such subroutines like building blocks. 

{\color{black}
Furthermore, the subroutines can be implemented as either function calls or as macros.
In case subroutines are implemented as function calls, in each module there is only one single copy of the code of each subroutine, which is executed with calls and returns.
This has the advantage of reducing the code size, and making a better use of instruction cache.
As an example, the subroutine implementing the {\tt gemm} loop is likely to be a hot codepath, as it is shared by all level 3 DLA routines. 
In case subroutines are implemented as macros, a copy of the code of each subroutine is inserted in each linear algebra kernel, avoiding the cost of calls and returns at the expense of larger code size and colder critical codepaths.
This can be advantageous in case of very small matrices, and in case few DLA routines are employed.
A compiler flag can be used to easily choose between the three cases: 1) all subroutines as function calls; 2) {\tt gemm} loop as function call and all other subroutines as macros; 3) all subroutines as macros.
}

\subsubsection{Handling of corner cases}

If the sizes of the result matrix are not exact multiples of the sizes of the optimal kernel or if the sub-matrices are not aligned to the top of a panel, the issue of handling corner cases arises.
In BLASFEO HP, this is handled by using a few
{\color{black}
computational kernels of fixed size and by maskinig out the uselessly computed elements while storing the result matrix.
For each kernel size, three variants (called nominal, variable size and generic) are implemented. 
}

Depending on the target architecture, a small number of kernels for each DLA routine is implemented (typically 1 to 3).
E.g., in the case of the Haswell architecture, the optimal {\tt dgemm} kernel has size $12\times 4$, but also the kernels of size $8\times 4$ and $4\times 4$ are implemented.
The smaller size is generally chosen as $p_s\times p_s$, i.e., such kernel processes one panel from $A$ and one panel from $B$.
{\color{black}
Even if it is possible to write DLA kernels tailored to explicitly handle sizes not multiple of $p_s$, they are generally not considered in the current BLASFEO HP implementation, as their performance improvement is too little compared to the additional amount of code that needs to be written.
For the Intel Haswell and Intel Sandy-Bridge architectures (both having $p_s=4$ in double precision), Section~\ref{sec:exp:small_gemm} compares the performance of the {\tt dgemm\_nn} routine in the cases of $4\times 4$ and $2\times 2$ minimum kernel sizes.
}

For each DLA kernel size, 
{\color{black}
three 
}
variants are implemented.
The \emph{nominal} variant computes a sub-matrix of the result matrix whose size is exactly equal to the kernel size.
These kernels give the smallest overhead, and are used to compute the interior of the result matrix.
The \emph{variable-size} variant internally computes a sub-matrix of size equal to the kernel size, but allows to store a smaller sub-matrix of the result matrix (masking out some rows and columns)
This allows to handle corner cases, at the expense of a slight overhead (to handle the extra logic required to decide what elements should be stored and where).
{\color{black}
Finally, the \emph{generalized} variant internally computes a sub-matrix of size equal to the kernel size and allows to store a smaller sub-matrix too, but in addition the sub-matrix is possibly non-aligned to the top of a panel (carrying over some elements to the following panel).
This allows to handle arbitrary sub-matrices, at the expense of some overhead.
}

The choice of having these 
{\color{black}
three}
variants of each DLA kernel is a trade-off giving reasonably good performance (as the matrix size increases, most of the computation is performed by the low-overhead nominal kernel) without requiring the explicit handling of many special cases.
The use of subroutines in BLASFEO allows the 
{\color{black} three
}
kernel variants to share all the code with the exception of the specialized store subroutines, avoiding code duplication.

\subsection{Order of outer loops} \label{sec:impl_det:out_loop}

The {\tt gemm} routine optimized for small matrices is implemented by means of two loops around the carefully optimized {\tt gemm} kernel,
{\color{black}
which consists of a single C-resident loop.
Due to the lack of cache blocking,
}
in case of a {\tt gemm} kernel where $m_r$ and $n_r$ are not equal, the order of the two outer loops has a big impact on the performance of the {\tt gemm} routine as the size of the factor matrices increase.

In BLASFEO HP, it is generally the case that $m_r>n_r$ (i.e. the {\tt gemm} kernel computes a sub-matrix of $C$ with more rows than columns) in architectures with SIMD instructions, since this reduces the number of shuffle or broadcast instructions. 
Therefore, in the {\tt gemm} kernel, the number of streamed panels from $A$ (i.e. $\sfrac{m_r}{p_s}$) is larger than the number of streamed panels from $B$ (i.e. $\sfrac{n_r}{p_s}$).
In order to minimize the memory movements between cache levels 
{\color{black}
and therefore not exceed L2 or L3 cache bandwidth,}
it is convenient to keep the $\sfrac{m_r}{p_s}$ panels from $A$ in L1 cache, while streaming the $\sfrac{n_r}{p_s}$ panels from $B$ from L2 or L3 cache (and therefore minimizing the amount of data that has to be loaded from L2 or L3 cache to compute the same amount of flops).
{\color{black}
This can be obtained by making the intermediate loop over the columns of the result matrix C (and therefore A-resident), and the outermost loop over the rows of the result matrix C (and therefore B-resident).
}

Ignoring cache associativity, as a rule of thumb this approach gives close to full performance in the computation of matrices with $k$ up to the value such that $m_r\cdot k + n_r\cdot k$ elements can fit in L1 cache at once.
In practice, this $k$ value is often in the range 200 to 400, large enough for most embedded optimization applications.
For larger values of $k$, performance can be recovered by adding blocking for different cache levels.
However, this is not of interest in the BLASFEO framework.


\subsection{BLAS and LAPACK implementation} \label{sec:impl_det:lapack}

If level 3 BLAS and LAPACK routines are implemented without packing, the {\tt gemm} kernel can not handle triangular factor matrices, triangular result matrices, factorizations, substitutions (i.e., solution of triangular system of equations) and inversions.
These operations require specialized routines.
Several approaches can be used in the implementation of these routines and in their use of the {\tt gemm} kernel.
	

\paragraph{Level 3 BLAS}
In optimized level 3 BLAS libraries, when packing is employed, it is possible to implement all level 3 BLAS routines (with the exception of {\tt trsm}, implementing substitutions) using the sole {\tt gemm} kernel and properly packing/padding routines~\cite{Goto2008a}.
The {\tt trsm} routine is an exception, since the downgrade part of the routine can be cast in terms of {\tt gemm} kernel, while the substitution part can not.
In~\cite{Zee2015}, two {\tt trsm} approaches are compared.
In one approach, the {\tt gemm} kernel is explicitly used for the downgrade, while another specialized routine 
takes care of the substitution part.
This approach has the advantage of requiring the design only of the {\tt gemm} kernel, but it has the drawback of larger overhead since there are two function calls and the result sub-matrix needs to be loaded and stored in memory twice.
In the other approach, the {\tt gemm} kernel and the specialized substitution routines are merged into a single {\tt trsm} kernel.
This requires the design of a specialized {\tt trsm} kernel, but it has lower overhead and therefore it gives better performance for small matrices.

In BLASFEO HP, the second approach is employed for the implementation of all level 3 BLAS-like routines, since it gives the best performance for small matrices.
Therefore, specialized kernels are designed, where the main loop is given by the {\tt gemm} assembly subroutine, while specialized assembly subroutines are called before and after this loop to take care of triangular matrices and substitutions.
The modularity of the BLASFEO HP assembly subroutine based approach implies that, once the {\tt gemm} kernel has been implemented, all other level 3 BLAS kernels can be easily coded at the cost of a little increase in the code size.

\paragraph{LAPACK}
LAPACK routines make use of BLAS routines, but in general not of BLAS kernels, since their interfaces are not standardized and therefore not exposed (the BLIS project is an exception, exposing also its lower level interface).
LAPACK contains both unblocked and blocked versions of all routines.
Unblocked versions make use of level 2 BLAS and elementary operations such as square roots and divisions.
They compute the result matrix one row or column at a time, and are usually employed for small matrices and as routines in blocked versions.
Blocked versions make use of level 3 BLAS and unblocked LAPACK routines for factorizations and substitutions (that are the matrix equivalent of square roots and divisions).
They compute the result matrix one sub-matrix at a time, and they rely on the underlying optimized BLAS routines to provide high-performance for large matrices.
In the context of embedded optimization, the main drawback of this approach is that it suffers from a considerable overhead (due to the many levels of routines), and the small-scale performance is therefore poor.

Some optimized BLAS libraries (as e.g. OpenBLAS) contain an optimized version of some of the key LAPACK routines (such as Cholesky and LU factorization, triangular matrix inversion, multiplication of two triangular matrices).
These routines are written making use of the optimized level 3 BLAS kernels (and not routines), and therefore exhibit a better performance for small matrices.
In particular, this allows the choice of a smaller threshold to switch to the blocked version of the algorithms, therefore casting more computations in the terms of the optimized level 3 BLAS kernels.

In BLASFEO HP, LAPACK-like routines are implemented in the same way as level 3 BLAS-like routines.
Namely, special kernels are written for the LAPACK-like routines as well. 
Therefore, there is not the equivalent of unblocked LAPACK routines, and the optimized kernels are used for all matrix sizes.
In other words, the block size of the blocked version of LAPACK routines is chosen to be equal to the {\tt gemm} kernel size, and the unblocked version of LAPACK routines is simply an assembly subroutine operating on a register-resident sub-matrix.
In case of small matrices, numerical tests show that this approach gives the best performance.

\subsection{Custom linear algebra routines} \label{sec:impl_det:custom}

The ability to customize linear algebra routines allows for further performance improvements, especially in the case of small matrices.
The RF and HP implementations of BLASFEO can take advantage of that, while the WR implementation can not, being simply a wrapper to standard BLAS and LAPACK.

\subsubsection{Inverse of diagonal in factorizations}

In algorithms for matrix factorizations (as e.g. Cholesky or LU), the inverse of the diagonal elements of the result matrix is computed as an intermediate step and generally discarded.
In BLASFEO RF and BLASFEO HP, in the matrix structure there is an additional pointer to memory, which points to an array of FP numbers large enough to hold any 1-dimensional sub-matrix.
In particular, this memory space can be used to save the inverse of the diagonal computed during factorizations.
The inverse of the diagonal can be employed in subsequent system solutions, removing the need to compute further FP divisions (that have considerably longer latency than multiplications).
The time saving is linear in the matrix size, and therefore it becomes negligible for large matrices.

\subsubsection{Fusing linear algebra routines}

As a motivating example, the convex equality constrained quadratic program
\begin{align*}
\min_x & \quad \tfrac 1 2 x^T H x + g^T x \\
s.t. & \quad A x + b = 0
\end{align*}
is considered, where the matrix $H$ is symmetric and positive definite. 
The Karush-Kuhn-Tucker (KKT) optimality conditions can be written as
\begin{equation*}
\begin{bmatrix} H & A^T \\ A & 0 \end{bmatrix}
\begin{bmatrix} x \\ \lambda \end{bmatrix} =
\begin{bmatrix} -g \\ -b \end{bmatrix}
\end{equation*}
which is a system of linear equations.
The KKT matrix is symmetric and indefinite.
One way to solve such a system is to use the range-space method, to compute the Schur complement of $H$ in the KKT  matrix, $-AH^{-1}A^T$, and to reduce the system to the so called normal equations
\begin{equation*}
- A H^{-1} A^T \lambda = - b + A H^{-1} g
\end{equation*}
If the matrix $A$ has full row rank, the Schur complement is a 
{\color{black}
negative
}
definite matrix and
{\color{black}
its opposite
}
can be Cholesky factorized to solve the normal equations.
The Cholesky factorization
{\color{black}
of the opposite
}
of the Schur complement can be computed efficiently as
\begin{equation*}
( AH^{-1}A^T )^{\sfrac{1}{2}} = ( A (L L^T)^{-1} A^T )^{\sfrac{1}{2}} = ( A L^{-T} L^{-1} A^T )^{\sfrac{1}{2}} = ( (A L^{-T}) (A L^{-T})^T )^{\sfrac{1}{2}}
\end{equation*}
(where the exponent $^{\sfrac{1}{2}}$ indicates the Cholesky factorization) by means of the following four calls to BLAS and LAPACK routines (using the BLASFEO convention of hard-coding the char arguments in the name):
\begin{description}
\item[potrf\_l] Cholesky factorization of the Hessian matrix $H = L L^T$ 
\item[trsm\_rltn] triangular system solution $A L^{-T}$ 
\item[syrk\_ln] symmetric matrix multiplication $(A L^{-T}) (A L^{-T})^T$
\item[potrf\_l] final Cholesky factorization of the Schur complement
\end{description}
The first and second linear algebra routine can be fused into a single custom one, as well as the third and fourth.

\paragraph{Stacking matrices} 
In the implementation of the Cholesky factorization, the routine {\tt trsm\_rltn} is employed to compute the off-diagonal blocks.
Therefore, it is natural to stack the $H$ and $A$ matrices as
\begin{equation*}
\begin{bmatrix} H \\ A \end{bmatrix}
\end{equation*}
and use a non-squared variant of the Cholesky factorization to fuse the routines {\tt potrf\_l} and {\tt trsm\_rltn} into a single one.
In the BLASFEO HP framework, this has the advantage that, depending on the matrix sizes, the stacked matrix may fit in a smaller number of panels than the total number of panels of the two original matrices. 
This is the case if $0 < {\rm rem}(m_H,p_s) + m_A \% p_s < p_s$, where $m_H$ and $m_A$ are the number of rows of the $H$ and $A$ matrices and ${\rm rem}(x,y)$ is the reminder of the division between $x$ and $y$.
Then, the stacked matrix can be processed using a smaller number of calls to {\tt trsm\_rltn} kernels (note that no new kernel needs to be coded).
This technique is especially advantageous in the case of small matrices, where the matrix sizes are not too large compared to the panel size $p_s$.

\paragraph{Concatenating updates/downdates}
In the implementation of the Cholesky factorization, the downdate of the sub-matrices is in the form of {\tt syrk\_ln} for the diagonal blocks and of {\tt gemm\_nt} for the off-diagonal blocks.
Therefore, it is natural to fuse the third and fourth routines in the previous example and to write a specialized kernel performing the update and downdate of each sub-matrix at once, without having to store and then load again the same data.
This reduces the overhead by increasing the amount of work that each linear algebra kernel performs, and therefore amortizing the cost to load and store the sub-matrices over a larger amount of rank-1 updates or downdates.
Also this technique is especially advantageous in the case of small matrices, where the rank of updates and downdates is typically lower.

In the BLASFEO HP framework, fused linear algebra kernels can be coded
very easily and at very little increase in code size, since no new
subroutines need to be coded, and the fused kernels are simply stacking
together calls to existing subroutines.

\subsection{Target architectures and operating systems} \label{sec:impl_det:target}

The current BLASFEO HP implementation supports the following target architectures and instruction set architectures (ISA).

\paragraph{X64\_INTEL\_HASWELL} x86\_64 architecture with FMA3 and AVX2 ISAs, code optimized for Intel Haswell and Intel Skylake microarchitectures.
The Haswell and Skylake cores have 256-bit wide execution units, and they can perform 2 fused-multiply-add (FMA) every clock cycle, with a latency of 5 (Haswell) or 4 (Skylake) clock cycles respectively.
The panel size $p_s$ is 4 in double precision and 8 in single precision.
The optimal kernel size is $12\times 4$ in double precision and $24\times 4$ in single precision.

\paragraph{X64\_INTEL\_SANDY\_BRIDGE} x86\_64 architecture with AVX ISA, code optimized for Intel Sandy Bridge microarchitecture.
The Sandy-Bridge core has 256-bit wide execution units, and it can perform 1 multiplication and 1 addition every clock cycle, with a latency of 5 and 3 clock cycles respectively.
The panel size $p_s$ is 4 in double precision and 8 in single precision.
The optimal kernel size is $8\times 4$ in double precision and $16\times 4$ in single precision.

\paragraph{X64\_INTEL\_CORE} x86\_64 architecture with SSE3 ISA, code optimized for Intel Core and Intel Nehalem microarchitectures.
The Core and Nehalem cores have 128-bit wide execution units, and they can perform 1 multiplication and 1 addition every clock cycle, with a latency of 4/5 (mul in single/double precision) and 3 (add) clock cycles.
The panel size $p_s$ is 4 in double precision and 4 in single precision.
The optimal kernel size is $4\times 4$ in double precision and $8\times 4$ in single precision.

\paragraph{X64\_AMD\_BULLDOZER} x86\_64 architecture with AVX and FMA3 ISAs, code optimized for AMD Bulldozer microarchitecture.
The Bulldozer core has 128-bit wide execution units, and it can perform 2 FMA every clock cycle, with a latency of 5-6 clock cycles.
Instructions operating on 256-bit vectors are executed as two instructions operating on 128-bit vectors; best performance is obtained explicitly targeting 128-bit vectors.
The panel size $p_s$ is 4 in double precision and 4 in single precision.
The optimal kernel size is $4\times 4$ in double precision and $12\times 4$ in single precision.

\paragraph{ARMV8A\_ARM\_CORTEX\_A57} ARMv8A architecture with NEONv2-VFPv4 ISAs, code optimized for ARM Cortex A57 core.
The Cortex A57 core has 128-bit wide execution units, and it can perform 1 FMA every clock cycle, with a latency of 10 (both double and single precision) clock cycles.
The panel size $p_s$ is 4 in double precision and 4 in single precision.
The optimal kernel size is $8\times 4$ in double precision and $8\times 8$ (or $16\times 4)$) in single precision.

\paragraph{ARMV7A\_ARM\_CORTEX\_A15} ARMv7A architecture with NEONv2-VFPv4 ISAs, code optimized for ARM Cortex A15 core.
The Cortex A15 core has 64-bit wide (double precision) and 128-bit wide (single precision) execution units, and it can perform 1 FMA every clock cycle, with a latency of 9 (double precision) and 10 (single precision) clock cycles.
The panel size $p_s$ is 4 in double precision and 4 in single precision.
The optimal kernel size is $4\times 4$ in double precision and $12\times 4$ in single precision.

\paragraph{GENERIC} Generic C code, targeting a typical RISC machine with 32 scalar FP registers.
The panel size $p_s$ is 4 in double precision and 4 in single precision.
The optimal kernel size is $4\times 4$ in double precision and $4\times 4$ in single precision.

\phantom{asdas}

The current BLASFEO HP implementation supports the following combinations of operating systems and compilers

\paragraph{LINUX} BLASFEO runs on Linux 32-bit (ARMv7A architecture) or 64-bit (x86\_64 and ARMv8A architectures) versions, with compilers {\tt gcc} or {\tt clang}.

\paragraph{MAC} BLASFEO runs on Mac 64-bit (x86\_64 architecture) version, with compilers {\tt gcc} and {\tt clang}.

\paragraph{WINDOWS} BLASFEO runs on Windows 64-bit (x86\_64 architecture) version, with compiler {\tt MinGW-w64}.

\phantom{asdas}

The x86\_64 assembly kernels in BLASFEO are written using the AT\&T syntax.
Therefore, they can not be employed directly in compilers that only accept the Intel syntax (e.g., Visual Studio).
{\color{black}
However, since BLASFEO HP assembly kernels do not have any external dependency, supported compilers like e.g. {\tt MinGW-w64} can be used to assembly the .S files into .o files, which can be included into a library using e.g. Visual Studio (together with the .c files compiled using the latter).
}

\section{Experiments}
\label{sec:exp}

This section contains  numerical experiments showing the performance of the proposed implementation approach.
Note that all software considered in the experiments is single-threaded.
Multi-thread experiments are outside the scope of the current paper, and object of future research.

{\color{black}

\subsection{{Small \tt dgemm} performance: comparison of library and JIT approaches} \label{sec:exp:small_gemm}

This section contains some numerical experiments comparing the performance of small {\tt dgemm} as implemented using library versus JIT approaches.
All tests are performed on recent Intel architectures (Ivy Bridge and Haswell).
The results are in figure~\ref{fig:exp:dgemm_small}.

\begin{figure}[!t]
\centering
\subfloat{\includegraphics[width=0.42\linewidth]{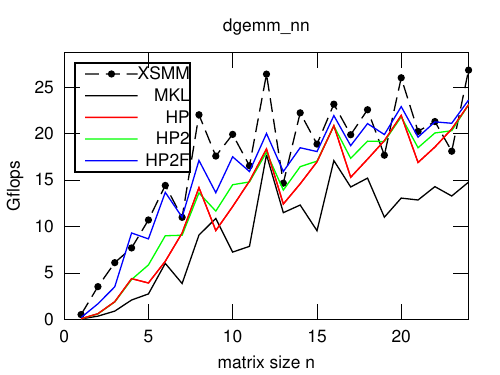} \label{fig:exp:sandybridge:dgemm_small}} 
\subfloat{\includegraphics[width=0.42\linewidth]{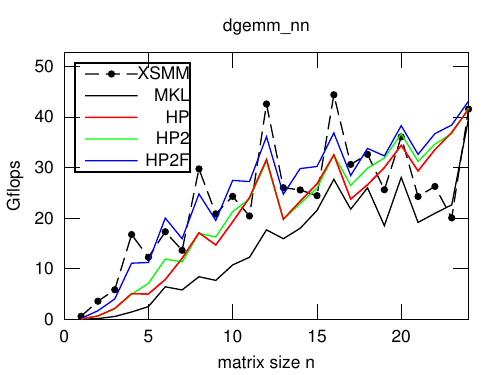} \label{fig:exp:haswell:dgemm_small}} \\
\caption{{\color{black}
Small-scale performance plot for selected implementation approaches for the {\tt dgemm\_nn} routine, on Intel Ivy-Bridge (left, maximum double-precision throughput 28.8 Gflops) and Intel Haswell (right, maximum double-precision throughput 52.8 Gflops).
HP2 refers to BLASFEO HP with $2\times 2$ minimum kernel size; HP2F refers to BLASFEO HP with $2\times 2$ minimum kernel size and with DLA routines tailored to fixed matrix sizes.
}
}
\label{fig:exp:dgemm_small}
\end{figure}

The proprietary BLAS and LAPACK implementation by Intel is part of MKL (Math Kernel Library, here tested in the version 2017.2.174, labelled {\tt MKL} in Figure~\ref{fig:exp:dgemm_small}).
MKL provides a library implementation of {\tt dgemm}, which can handle any matrix size.
The linking flag {\tt MKL\_DIRECT\_CALL\_SEQ} enables the use of a fast path for selected DLA routines (comprising {\tt gemm}), with the aim of improving performance in case of small matrices.

Intel also supports the open-source project LIBXSMM \cite{Heinecke2016}, which provides a JIT framework to generate selected routines optimized for small-scale performance.
The tested version is master 1.8.1-1204, labelled {\tt XSMM} in Figure~\ref{fig:exp:dgemm_small}.
Out of level 3 BLAS and LAPACK routines, LIBXSMM currently provides highly optimized implementations only of the NN variant of {\tt gemm}, with a limited set of options for the routine arguments (covering the most common use cases).
For small matrices, the performance of the {\tt gemm}  routine provided by LIBXSMM outperforms other JIT implementations \cite{Heinecke2016}.
In our experiments, only the C interface of LIBXSMM is considered.
LIBXSMM employs a memory buffer to store {\tt gemm} kernels optimized for specific operation descriptors (comprising e.g. the operation sizes {\tt(m, n, k)} and the leading dimension of matrices).
The first time a new descriptor is encountered, the execution falls back to a (statically generated) library version implemented using SSE4.2, while the JIT framework generates the optimized kernel.
Once the JIT kernel generation is completed, subsequent calls with the same descriptor employ the optimized kernel.
For best performance, LIBXSMM provides the possibility to explicitly trigger the JIT generation of a kernel, which is dispatched using a function pointer and therefore it can be called directly.

As described in previous sections, BLASFEO is implemented as a library.
Currently BLASFEO HP (labelled {\tt HP} in Figure~\ref{fig:exp:dgemm_small}) is implemented using few kernel sizes, with the minimum kernel size for level 3 DLA routines equal to $4\times 4$.
This choice is a trade-off between performance on 'odd' sizes (i.e. sizes not multiple of the minimum kernel size) on one hand, and required number of kernels and code size on the other hand.
E.g. in the case of the Intel Sandy-Bridge architecture, the optimal kernel size is $8\times 4$, but also the kernels of size $12\times 4$ and $4\times 4$ are implemented.
The variable-size variant of each kernel makes use of mask store instructions, therefore covering any matrix size between 1 and 12 rows and between 1 and 4 columns.
Naturally, if the matrix size is not multiple of 4, unnecessary FP computations are performed. 
For the tested matrix sizes, BLASFEO HP is generally faster than MKL with small-size fast path (and therefore it is the fastest library approach), but slower than LIBXSMM.

In this section, we investigate the effect of reducing the minimum kernel size to $2\times 2$ (and therefore considering kernel sizes multiple of 2) in the high-performance version of BLASFEO (labelled {\tt HP2} in Figure~\ref{fig:exp:dgemm_small}).
This greatly increases the number of possible kernels (e.g. also $6\times 8$, $8\times 6$, $6\times 6$, $10\times 4$, $6\times 4$, $2\times 4$, $10\times 2$, $8\times 2$, $6\times 2$, $4\times 2$, $2\times 2$ can be kernel sizes).
High-performance implementation of these kernels (that fully exploit vector units, even if the kernel size is not a multiple of the SIMD width 4) allows to increase performance for 'odd' sizes.
However, as the matrix size increases (and therefore the computational performance of {\tt gemm} increases), the performance improvement relative to the {\tt HP} version shrinks to below 10\%.
And for very small matrix sizes, up to 4, this does not seem to improve performance, which remains much lower than LIBXSMM.

As a final experiment, we investigate the performance of HP2 routines specialized for matrices of fixed sizes, described by the triplet {\tt(m, n, k)}.
This version is labelled {\tt HP2F} in Figure~\ref{fig:exp:dgemm_small}.
The DLA kernels are unchanged compared to the HP2 version, but the two loops and the logic around them is removed, replaced by the exact sequence of needed kernel calls.
In this experiment the exact sequence is hard coded, but it could be very easily and cheaply generated using a JIT approach.
For very small sizes, this gives over 2.5 times speedup compared to HP or HP2.
The performance gets much closer to the one of LIBXSMM, and at times exceeds it, especially as the matix size increases.
The performance advantage still present in LIBXSMM for e.g. size 12 is mainly due the fact that, for sizes smaller than 16, the loops in LIBXSMM are fully unrolled, while in the BLAFEO kernels the innermost loop size is still fixed to 4, since kernels are untouched.

As a conclusion to this section, in case of very small matrix sizes, the JIT approach used in LIBXSMM is inherently superior to the library approach used in BLASFEO and MKL.
Out of the library approaches, BLASFEO provides better performance than MKL, even more if the minimum kernel size in BLASFEO is reduced to $2\times 2$.

LIBXSMM is not further considered in this paper, since it currently provides only {\tt gemm} and therefore it is of limited interest in embedded optimization, where other routines like factorizations and triangular system solutions are needed.

}

\subsection{BLAS- and LAPACK-like routines} \label{sec:exp:blas}

This section contains the result of numerical experiments on the performance of key linear algebra routines.
%
%
Section \ref{sec:exp:blas:choice} presents many approaches for the implementation of the linear algebra routines, and motivates the choice of some of them for the tests in the following sections.
The experiments in this section focus on small-scale performance, while the scalability with the matrix size is investigated in the following sections.
In Section \ref{sec:exp:blas:haswell} there are performance plots for the Intel Haswell processor, which is an example of a high-performance architecture implementing the latest ISAs.
In Section \ref{sec:exp:blas:sandybridge} there are performance plots for the Intel Ivy-Bridge processor, which is an example of a high-performance architecture with slightly older ISA.
In Sections \ref{sec:exp:blas:cortex_a57} and \ref{sec:exp:blas:cortex_a15} there are performance plots for the ARM Cortex A57 and A15 processors, which are examples of relatively low-power architectures that require more careful implementation.

For the Intel Haswell architecture, the computational performance of many BLAS- and LAPACK-like routines is reported.
For the sake of space, only the computational performance of the {\tt gemm\_nt} and {\tt potrf\_l} routines is reported for the remaining architectures.

\subsubsection{Choice of alternative approaches} \label{sec:exp:blas:choice}

\begin{figure}[!t]
\centering
\subfloat{\includegraphics[width=0.42\linewidth]{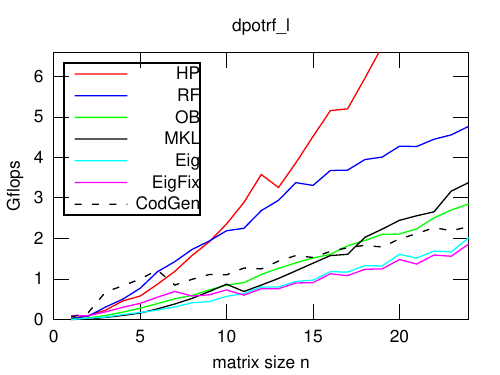} \label{fig:exp:haswell:dpotrf_small}} 
\subfloat{\includegraphics[width=0.42\linewidth]{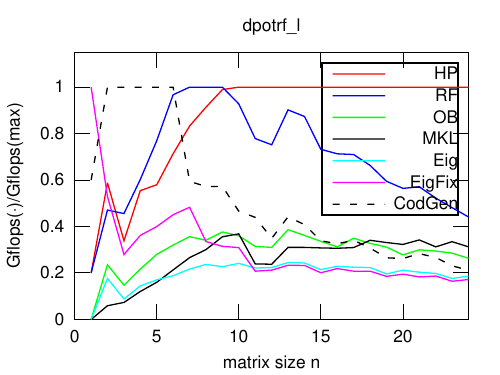} \label{fig:exp:haswell:dpotrf_small_norm}} \\
\subfloat{\includegraphics[width=0.42\linewidth]{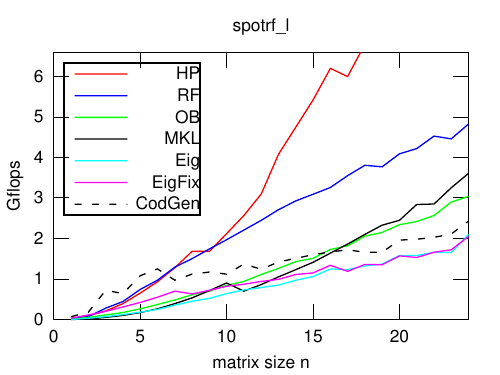} \label{fig:exp:haswell:spotrf_small}} 
\subfloat{\includegraphics[width=0.42\linewidth]{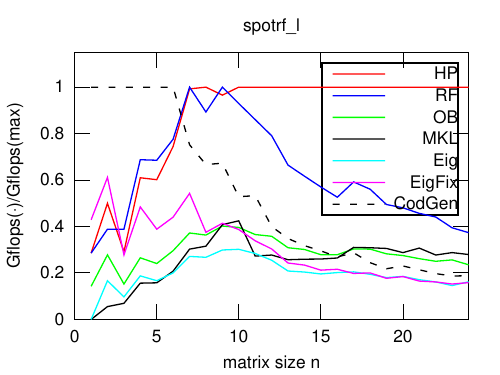} \label{fig:exp:haswell:spotrf_small_norm}} \\
\caption{Small-scale performance plot for all considered implementation approaches for the Cholesky factorization routine {\tt portf\_l}.}
\label{fig:exp:haswell:potrf_small}
\end{figure}

This section tests many approaches for the implementation of the Cholesky factorization, and compares them for small matrices of size $n$ up to 24.
Figure~\ref{fig:exp:haswell:potrf_small} show the computational performance of the considered approaches.

The test machine is a laptop equipped with an Intel Core i7 4800MQ (Intel Haswell architecture), which under AVX-heavy loads runs at 3.3 GHz, giving a maximum throughput in double (single) precision of 52.8 (105.6) Gflops.
The OS is Linux with kernel 4.4.
All code is compiled with {\tt gcc 5.4}.

Figure~\ref{fig:intro} shows that for small matrices, OpenBLAS outperforms the other open-source alternatives, namely ATLAS and BLIS.
Therefore, the latter libraries are not further considered.

The following implementations (coming from both the high-performance computing and the embedded optimization communities) are considered:
\begin{description}
\item[HP] - BLASFEO HP: the high-performance implementation of BLASFEO is the main contribution of this paper.
The linear algebra kernels are optimized for the Intel Haswell architecture.
Compiler flags: {\tt -O2 -mavx2 -mfma}.
\item[RF] - BLASFEO RF: the reference version of BLASFEO has small code size and easy embeddability as main aims.
Compiler flags: {\tt -O2 -mavx2 -mfma}.
\item[OB] - BLASFEO WR with OpenBLAS: OpenBLAS (version 0.2.19) is a highly optimized library with hand-crafted assembly kernels.
It is probably the best open-source alternative.
The DLA kernels are optimized for the Intel Haswell architecture.
Compiled disabling multi-threading, because this gives the best performance for single-threaded code.
\item[MKL]: - BLASFEO WR with Intel's Math Kernel Library (MKL, version 2017.2.174) is the best proprietary alternative on Intel processors.
Single-thread version.
The linking flag {\tt MKL\_DIRECT\_CALL\_SEQ} is used to reduce overhead.
\item[Eig] - Eigen: Eigen is advertised as offering very good performance and portability using C++ template headers.
Compiler flags: {\tt -O3 -mavx2 -mfma}.
The option to export a BLAS library does not work in the current version, so Eigen is not employed in all tests.
EIGEN\_NO\_DEBUG mode is chosen to reduce overhead.
\item[EigFix] - Eigen fix size: in Eigen, it is possible to hard code the size of matrices, allowing one to auto generate optimized code.
Compiler flags: {\tt -O3 -mavx2 -mfma}. 
EIGEN\_NO\_DEBUG mode is chosen to reduce overhead.
\item[CodGen] - Code-generated triple-loop: this is a C coded triple-loop version of the Cholesky factorization (that is, a C translation of the LAPACK unpacked routine {\tt potf2}), where the size of the matrices is fixed at compile time.
Compiler flags: {\tt -O3 -mavx2 -mfma -funroll-loops}. 
\end{description}

As a first note, for such small matrices the difference in performance between single and double precision is small, as the sequential parts of the Cholesky factorization algorithm (and especially divisions and square roots) dominate the vectorizable parts.

For $n$ up to roughly 6, the code-generated triple-loop version is the fastest, but it is quickly outperformed by BLASFEO RF and BLASFEO HP as $n$ increases.
BLASFEO RF does not require recompilation for each value of $n$ and its performance scales much better than code-generated triple-loop.
Therefore it is the overall best choice for small matrices.
For $n$ larger than roughly 10, the performance of BLASFEO HP quickly increases: for $n=24$, it exceeds 10 Gflops in both double and single precision.

Eigen with fixed code sizes performs better than BLASFEO HP for sizes up to 2, but it performs worse than code-generated triple-loop.
In Eigen, the option to fix the matrix sizes improves performance only for very small matrices, but for matrices larger than about 15 it decreases performance. 
All other alternatives show a rather low performance, as they need much larger $n$ for the performance to increase, with OpenBLAS and MKL performing slightly better than Eigen.

\paragraph{Remarks}
As a conclusion to this first set of tests, code generation approaches (code-generated triple-loop and Eigen with fixed matrix sizes) outperform the approaches proposed in BLASFEO only for very small matrices, of size up to 6, but they have the burden of having code tailored for a specific matrix size.
The approach used in BLASFEO RF gives better scalability with the problem size, and a portable and simple code.
Therefore code generation approaches are not further considered in the remaining tests.
The performance of BLASFEO HP increases quickly as soon as divisions and square roots are not the bottleneck and vectorization pays off.


\subsubsection{Intel Haswell} \label{sec:exp:blas:haswell}

This section contains performance plots for some linear algebra routines on the Intel Haswell architecture that targets the most recent ISAs in x86\_64 laptops/workstations.
The matrix size ranges in steps of 4 from 4 up to 300, which is large enough for most embedded optimization applications.

Haswell is a deeply out-of-order architecture, performing aggressive hardware prefetch.
It is relatively easy to write {\tt gemm} kernels giving high-performance, provided that at least 10 accumulation registers are employed.
The Haswell core can perform 2 256-bit wide FP fused-multiplication-accumulate every clock cycle, giving a throughput of 16 and 32 flops per cycle in double and single precision respectively.

In the implementation of BLASFEO HP, the panel size $p_s$ is 4 in double precision and 8 in single precision.
The optimal BLASFEO HP {\tt gemm} kernel size is $12\times 4$ in double precision and $24\times 4$ in single precision.
Hardware prefetch can detect the streaming of data along panels.

The test processor is the Intel Core i7 4800M (Haswell), running at 3.3 GHz when the 256-bit execution units are employed (3.7 GHz when they are disabled).
The memory is 8 GB of DDR3L-1600 RAM in dual-channel configuration, giving a bandwidth of 25.6 GB/s.

The performance of many DLA routines is reported for the Haswell architecture.
Two {\tt gemm} variants are tested, in both single and double precision.
These two variants are the backbone of most other DLA routines.
A second and a third set of tests investigate the performance of BLAS and LAPACK routines respectively.

\paragraph{gemm\_nt}
The {\tt gemm\_nt} is the general matrix-matrix multiplication with options 'nontransposed' and 'transposed'.
The {\tt gemm\_nt} is the optimal {\tt gemm} variant in BLASFEO HP, as it optimally streams both $A$ and $B$, i.e. along panels.
The {\tt gemm\_nt} subroutine is used in many kernels such as symmetric matrix-matrix multiplication and Cholesky factorization.

\paragraph{gemm\_nn}
The {\tt gemm\_nn} is the general matrix-matrix multiplication with options 'nontransposed' and 'nontransposed'.
In BLASFEO HP, it streams $A$ in an optimal way, along panels, while $B$ is streamed across panels, requiring software prefetch to hint the processor about this more complex memory access.
The {\tt gemm\_nt} subroutine is used in many kernels such as LU factorization.

\begin{figure}
\centering
\subfloat{\includegraphics[width=0.42\linewidth]{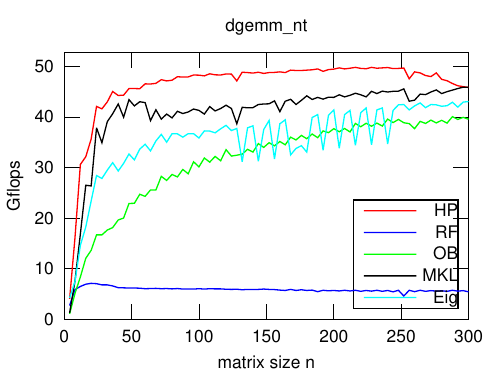} \label{fig:exp:haswell:dgemm_nt}} 
\subfloat{\includegraphics[width=0.42\linewidth]{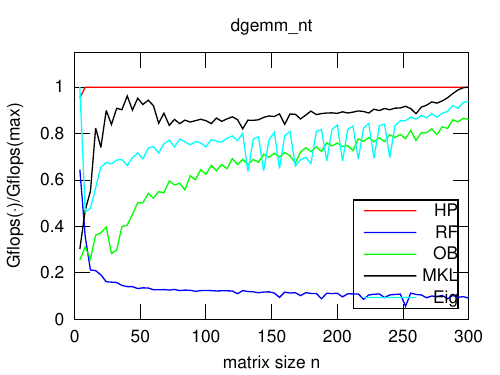} \label{fig:exp:haswell:dgemm_nt_norm}} \\
\subfloat{\includegraphics[width=0.42\linewidth]{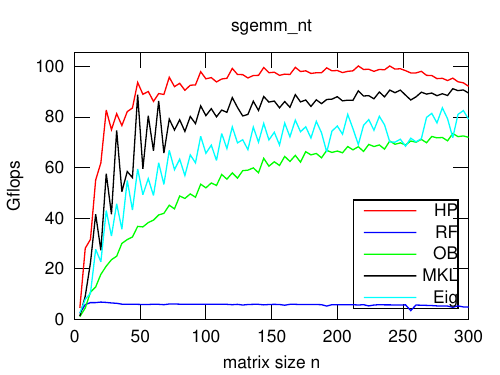} \label{fig:exp:haswell:sgemm_nt}} 
\subfloat{\includegraphics[width=0.42\linewidth]{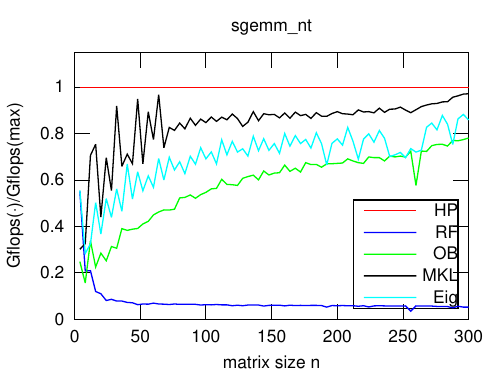} \label{fig:exp:haswell:sgemm_nt_norm}} \\
\subfloat{\includegraphics[width=0.42\linewidth]{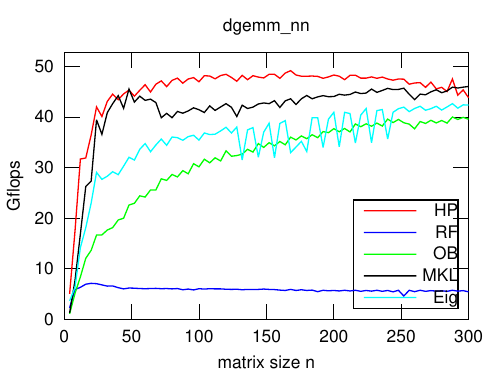} \label{fig:exp:haswell:dgemm_nn}} 
\subfloat{\includegraphics[width=0.42\linewidth]{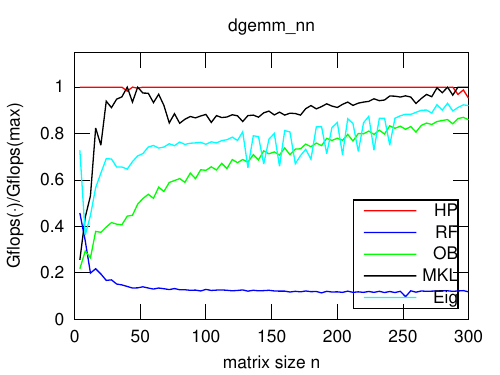} \label{fig:exp:haswell:dgemm_nn_norm}} \\
\subfloat{\includegraphics[width=0.42\linewidth]{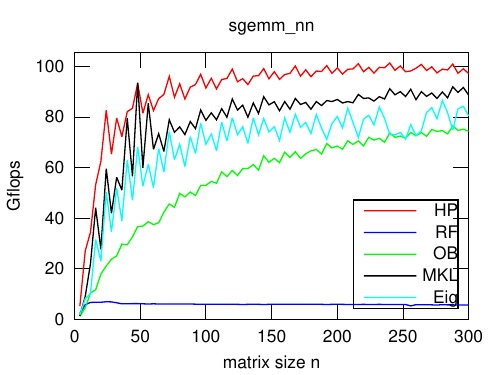} \label{fig:exp:haswell:sgemm_nn}} 
\subfloat{\includegraphics[width=0.42\linewidth]{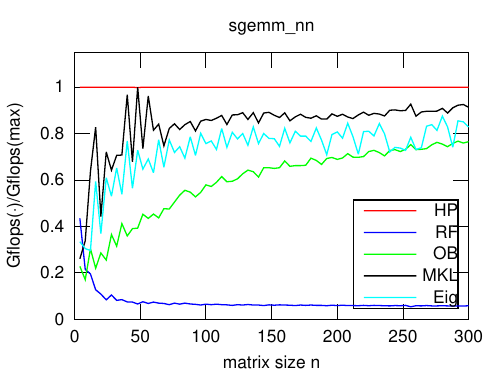} \label{fig:exp:haswell:sgemm_nn_norm}} \\
\caption{Performance of {\tt gemm\_nt} and {\tt gemm\_nn} on Intel Haswell.}
\label{}
\end{figure}

\paragraph{syrk\_ln}
The {\tt syrk\_ln} is the symmetric matrix-matrix multiplication, with options 'lower' and 'nontransposed'.
In BLASFEO, the left and right factor can be different matrices.
In BLASFEO HP, the {\tt syrk\_ln} routine is implemented using two kind of kernels, the {\tt gemm\_nt} (for the off-diagonal blocks) and {\tt syrk\_ln} (for the diagonal blocks), both implemented using the {\tt gemm\_nt} subroutine.

\paragraph{trmm\_rlnn}
The {\tt trmm\_rlnn} is the triangular matrix-matrix multiplication, with options 'right', 'lower', 'nontransposed', 'not-unit'.
In BLASFEO HP, it is implemented using a specialized kernel, which employs the {\tt gemm\_nn} subroutine.

\paragraph{trsm\_rltn}
The {\tt trsm\_rlnn} is the triangular system solve with matrix right-hand-side, with options 'right', 'lower', 'transposed', 'not-unit'.
In BLASFEO HP, it is implemented using a specialized kernel (employed also in the {\tt potrf\_l} routine), which employes the {\tt gemm\_nt} subroutine.

\begin{figure}
\centering
\subfloat{\includegraphics[width=0.42\linewidth]{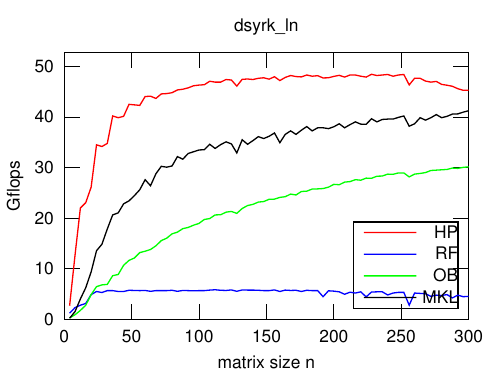} \label{fig:exp:haswell:dsyrk_ln}} 
\subfloat{\includegraphics[width=0.42\linewidth]{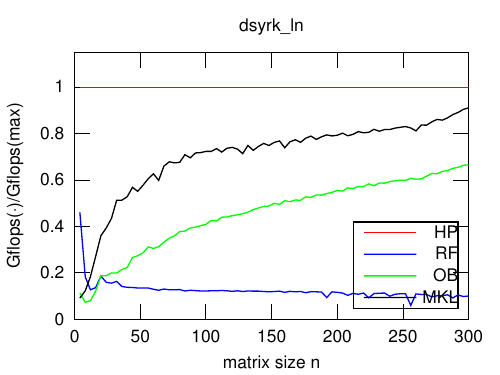} \label{fig:exp:haswell:dsyrk_ln_norm}} \\
\subfloat{\includegraphics[width=0.42\linewidth]{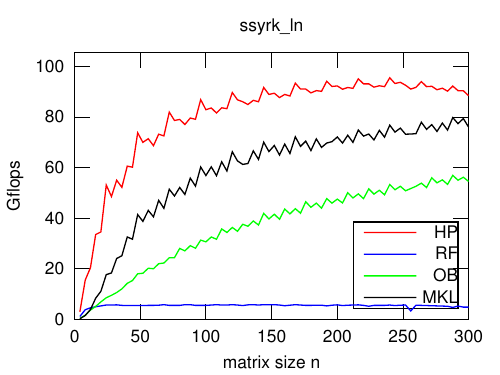} \label{fig:exp:haswell:ssyrk_ln}} 
\subfloat{\includegraphics[width=0.42\linewidth]{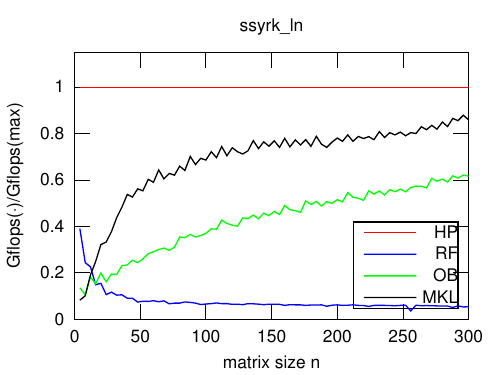} \label{fig:exp:haswell:ssyrk_ln_norm}} \\
\subfloat{\includegraphics[width=0.42\linewidth]{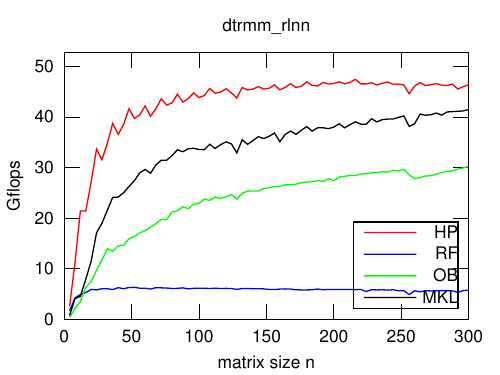} \label{fig:exp:haswell:dtrmm_rlnn}} 
\subfloat{\includegraphics[width=0.42\linewidth]{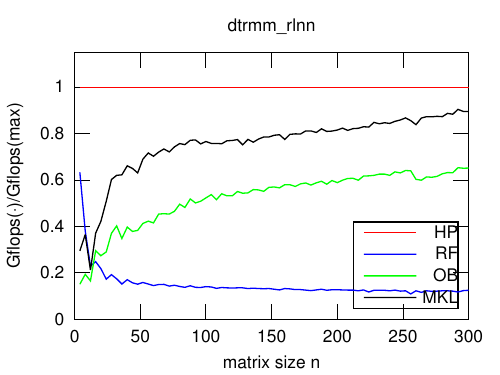} \label{fig:exp:haswell:dtrmm_rlnn_norm}} \\
\subfloat{\includegraphics[width=0.42\linewidth]{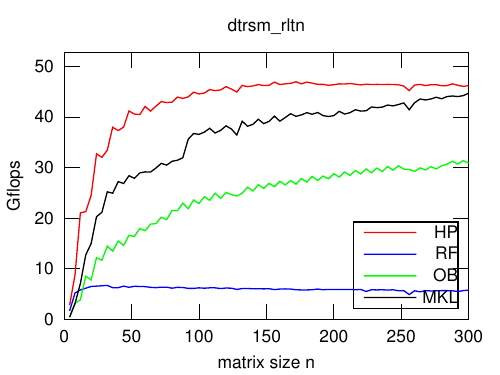} \label{fig:exp:haswell:dtrsm_rltn}} 
\subfloat{\includegraphics[width=0.42\linewidth]{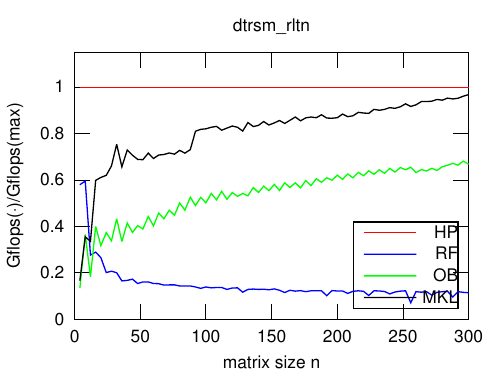} \label{fig:exp:haswell:dtrsm_rltn_norm}} \\
\caption{Performance of BLAS routines on Intel Haswell.}
\label{}
\end{figure}

\paragraph{potrf\_l}
The {\tt potrf\_l} is the routine computing the lower triangular Cholesky factorization, with option 'lower'.
This factorization is widely employed in embedded optimization.
In BLASFEO HP, the {\tt potrf\_l} routine is implemented using two kernels, the {\tt trsm\_rltn} (for the off-diagonal blocks) and {\tt potrf\_l} (for the diagonal blocks), both implemented using the {\tt gemm\_nt} subroutine.

\paragraph{getrf}
The {\tt getrf} is the routine computing the LU factorization with partial pivoting, which is part of LAPACK.
In BLASFEO HP, the kernels employed in the {\tt getrf} routine make use of the {\tt gemm\_nn} subroutine.

\paragraph{gelqf}
The {\tt gelqf} is the routine computing the LQ factorization, which is part of LAPACK.
This factorization is commonly employed in embedded optimization.
In BLASFEO HP, the {\tt dgelqf} routine is implemented using a blocked Householder LQ factorization with block size 4 for matrix sizes $n<128$, and with block size 12 for matrix sizes $n\geq 128$.
The routine employes the {\tt dgemm\_nt} kernel and the {\tt dger4} and {\tt dger12} kernels (performing a rank-4 and rank-12 update of a general matrix, respectively).

\begin{figure}
\centering
\subfloat{\includegraphics[width=0.42\linewidth]{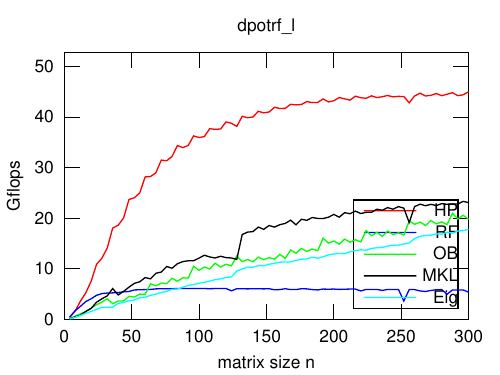} \label{fig:exp:haswell:dpotrf}} 
\subfloat{\includegraphics[width=0.42\linewidth]{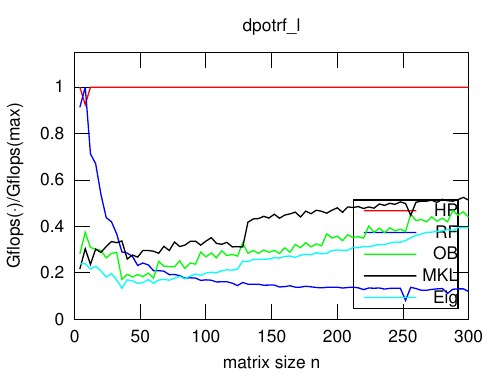} \label{fig:exp:haswell:dpotrf_norm}} \\
\subfloat{\includegraphics[width=0.42\linewidth]{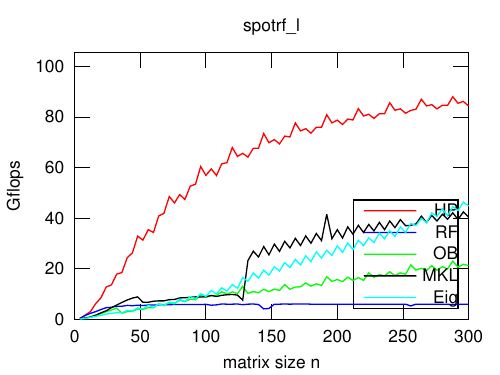} \label{fig:exp:haswell:spotrf}} 
\subfloat{\includegraphics[width=0.42\linewidth]{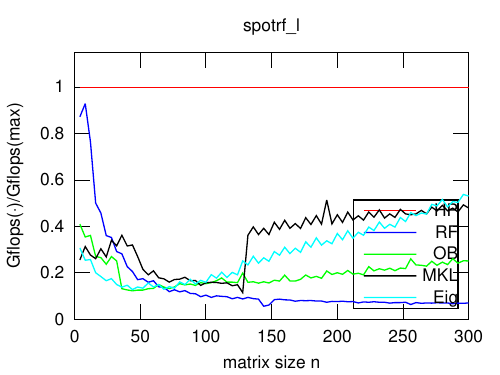} \label{fig:exp:haswell:spotrf_norm}} \\
\subfloat{\includegraphics[width=0.42\linewidth]{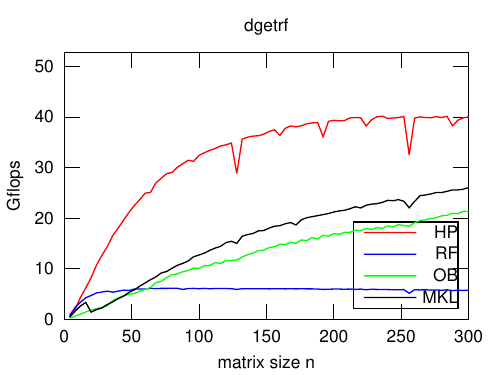} \label{fig:exp:haswell:dgetrf}} 
\subfloat{\includegraphics[width=0.42\linewidth]{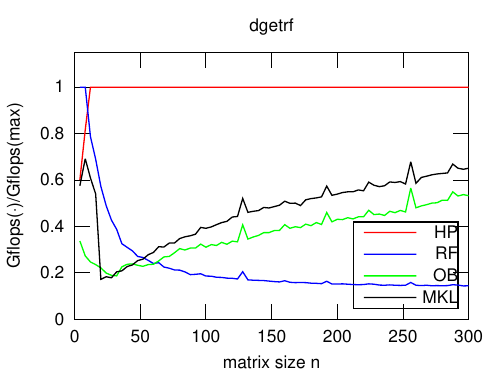} \label{fig:exp:haswell:dgetrf_norm}} \\
\subfloat{\includegraphics[width=0.42\linewidth]{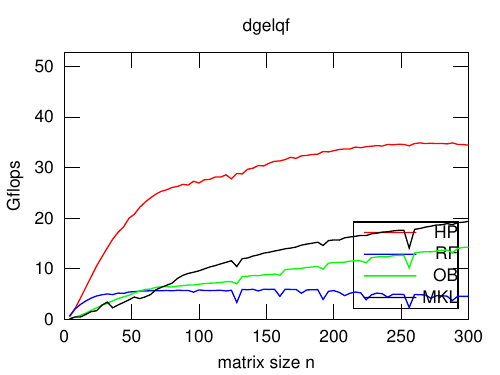} \label{fig:exp:haswell:dgelqf}} 
\subfloat{\includegraphics[width=0.42\linewidth]{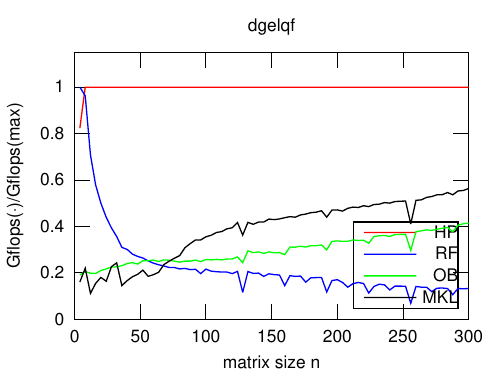} \label{fig:exp:haswell:dgelqf_norm}} \\
\caption{Performance of LAPACK routines on Intel Haswell.}
\label{}
\end{figure}

\paragraph{Remarks}
All experiments show that BLASFEO HP is clearly the best choice for the matrix sizes of interest, i.e. of sizes up to 300.
In particular, for matrices up to about 100, the speedup is of at least 20-30\% with respect to the best available BLAS implementation (usually provided by MKL), and in the order of 2-3 times with respect to the best available LAPACK implementation (again, usually provided by MKL).
%
BLASFEO RF performs well for very small matrices, in which case it is able to outperform optimized BLAS and especially LAPACK implementations.
In the {\tt gemm} tests, Eigen performs particularly well for matrices of size $4\times 4$, suggesting that this case is probably handled with a dedicated implementation.


\subsubsection{Intel Ivy-Bridge} \label{sec:exp:blas:sandybridge}

The Intel Ivy-Bridge is a high-performance architecture from a few years ago.
It is based on the Sandy-Bridge microarchitecture.
It supports the AVX ISA, but it lacks the AVX2 and FMA3 ISAs supported by the more recent Intel Haswell architecture.
Execution units are 256-bit wide, and they can process 4 doubles or 8 floats at a time.
The Ivy-Bridge core can perform one 256-bit wide FP multiplication and one 256-bit wide FP addition at every clock cycle, giving a throughput of 8 and 16 flops per cycle in double and single precision respectively.

The target architecture used in BLASFEO HP is Sandy-Bridge.
In the implementation of BLASFEO HP, the panel size $p_s$ is 4 in double precision and 8 in single precision.
The optimal BLASFEO HP {\tt gemm} kernel size is $8\times 4$ in double precision and $16\times 4$ in single precision.
Hardware prefetch can detect the streaming of data along panels.

The test processor is the Intel Core i7 3520M (Ivy-Bridge), running at 3.6 GHz during all tests.
The maximum throughput in double (single) precision is 28.8 (57.6) Gflops.
The memory is 8 GB of DDR3-1600 RAM in dual-channel configuration, giving a bandwidth of 25.6 GB/s.

In the case of the Intel Ivy-Bridge, only the {\tt gemm\_nt} and {\tt potrf\_l} routines are tested.
DLA libraries targeting the Sandy-Bridge microarchitecture should be in a mature state nowadays: this makes the comparison particularly meaningful.

\paragraph{gemm\_nt}
BLASFEO HP can achieve up to 95\% of the peak throughput.
It performs better than all alternatives in the matrix sizes of interest, followed by MKL.
For matrix sizes up to 50, the speedup compared to MKL is in the range of 20-30\%, which reduces to 5-10\% for matrix sizes up to 300.
OpenBLAS performs a little worse, especially in single precision, where it appears to employ a less performing kernel.
Eigen shows a quite solid performance in single precision, but a rather erratic performance in double precision.
As expected, BLASFEO RF is competitive only for very small matrices.

\paragraph{potrf\_l}
The results for the Cholesky factorization routine on Intel Ivy-Bridge are similar to the ones on Intel Haswell.
Namely, BLASFEO HP is about 2-3 times faster than OpenBLAS and MKL.
BLASFEO RF performs well for small matrices, outperforming optimized implementations for dimensions up to about 30. 

\begin{figure}
\centering
\subfloat{\includegraphics[width=0.42\linewidth]{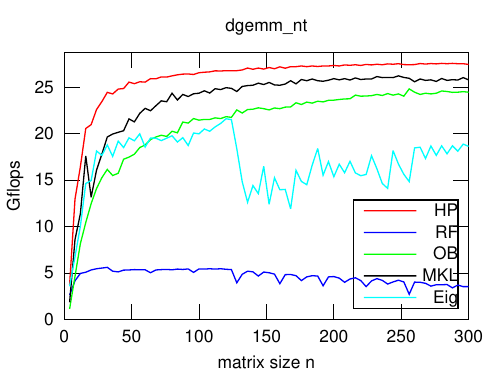} \label{fig:exp:sandybridge:dgemm_nt}} 
\subfloat{\includegraphics[width=0.42\linewidth]{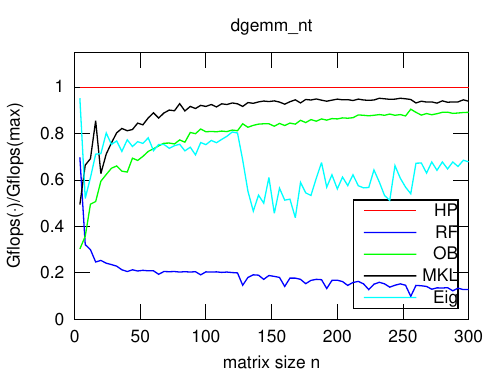} \label{fig:exp:sandybridge:dgemm_nt_norm}} \\
\subfloat{\includegraphics[width=0.42\linewidth]{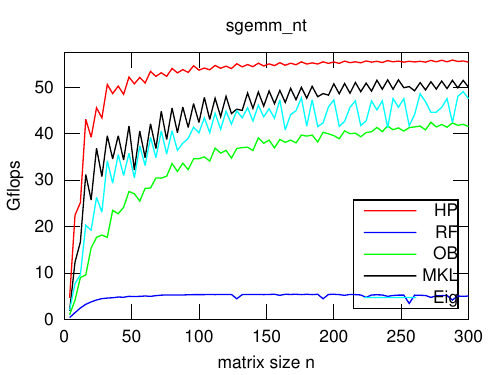} \label{fig:exp:sandybridge:sgemm_nt}} 
\subfloat{\includegraphics[width=0.42\linewidth]{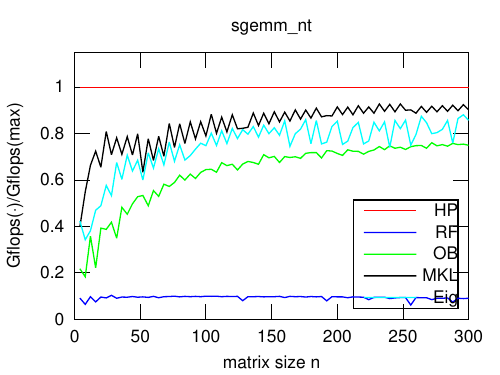} \label{fig:exp:sandybridge:sgemm_nt_norm}} \\
\subfloat{\includegraphics[width=0.42\linewidth]{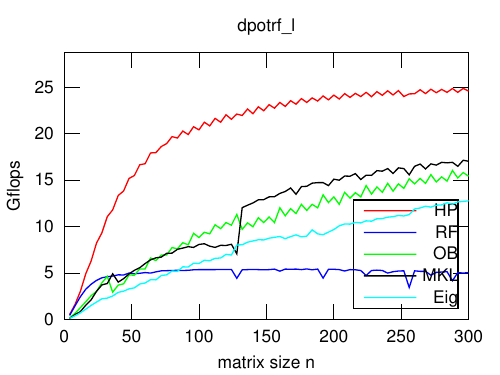} \label{fig:exp:sandybridge:dpotrf}} 
\subfloat{\includegraphics[width=0.42\linewidth]{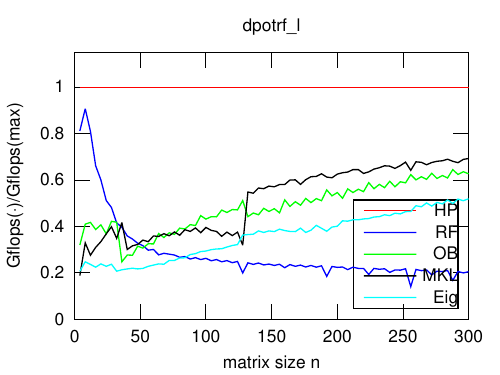} \label{fig:exp:sandybridge:dpotrf_norm}} \\
\subfloat{\includegraphics[width=0.42\linewidth]{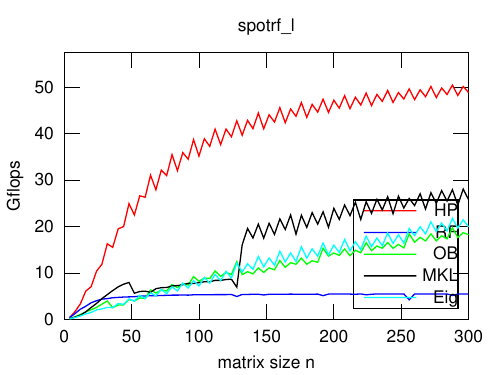} \label{fig:exp:sandybridge:spotrf}} 
\subfloat{\includegraphics[width=0.42\linewidth]{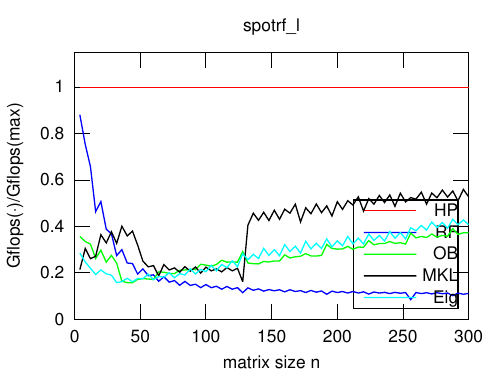} \label{fig:exp:sandybridge:spotrf_norm}} \\
\caption{Performance of {\tt gemm\_nt} and {\tt potrf\_l} on Intel Ivy-Bridge.}
\label{}
\end{figure}


\subsubsection{ARM Cortex A57} \label{sec:exp:blas:cortex_a57}

The ARM Cortex A57 is a relatively low-power architecture, and it is the 64-bit successor of the ARM Cortex A15.
It is a 3-way superscalar architecture with out-of-order execution.
The NEON ISA in the ARMv8A architecture supports vectorization in both single and double precision, with 4- and 2-wide vectors respectively.
The Cortex A57 core can perform a 128-bit wide FP fused-multiplication-accumulate at every clock cycle, giving a throughput of 8 and 4 flops per cycle in single and double precision respectively.
{\color{black}
Each core has 48 KB 3-way associative instruction L1 cache and 32 KB 2-way associative data L1 cache.
All cores share a 16-way associative unified L2 cache.
The cache line size is 64 bytes.
}

In the implementation of BLASFEO HP, the panel size $p_s$ is 4 in both double and single precision.
The optimal BLASFEO HP {\tt gemm} kernel size is $8\times 4$ in double precision and $8\times 8$ (or $16\times 4$) in single precision.
Software prefetch is employed for both the left and the right factors, slightly improving performance.

The test processor is the NVIDIA Tegra TX1 SoC (running at 2.15 GHz during all tests) in the Shield TV.
The memory interface is 64-bit LPDDR4-3200 giving 25.6 GB/s of bandwidth.
The amount of memory is 3 GB.

In the case of the ARM Cortex A57, only the {\tt gemm\_nt}
{\color{black}
and {\tt potrf\_l} routines are
}
tested.

\paragraph{gemm\_nt}
Also for this architecture, BLASFEO HP gives the best performance, reaching 90\% and 93\% of full throughput in double and single precision respectively.
The performance is steady and does not show negative spikes.
The performance of OpenBLAS is generally good, and its {\tt gemm} kernels give similar performance as the BLASFEO HP ones.
However, there are negative spikes at certain matrix sizes.
That could be due to the use of the column-major matrix format for an architecture with small cache associativity.
BLASFEO RF performs rather well for matrices fitting in L1 cache, but the performance deteriorates for larger matrices and it shows negative peaks due to cache associativity.

{\color{black}
\paragraph{potrf\_l}
In case of the Cholesky factorization {\tt potrf}, BLASFEO HP gives the best performance for nearly all tested matrix sizes, with the exception of very small sizes where BLASFEO RF may be the best choice.
Compared to OpenBLAS, BLASFEO HP gives a speed up of about 2 to 3 times for matrices of size up to 100.
}

\begin{figure}
\centering
\subfloat{\includegraphics[width=0.42\linewidth]{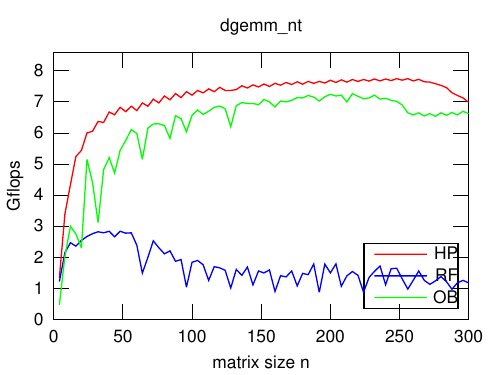} \label{fig:exp:cortex_a57:dgemm_nt}} 
\subfloat{\includegraphics[width=0.42\linewidth]{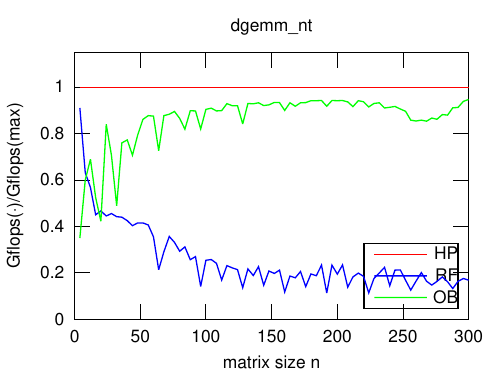} \label{fig:exp:cortex_a57:dgemm_nt_norm}} \\
\subfloat{\includegraphics[width=0.42\linewidth]{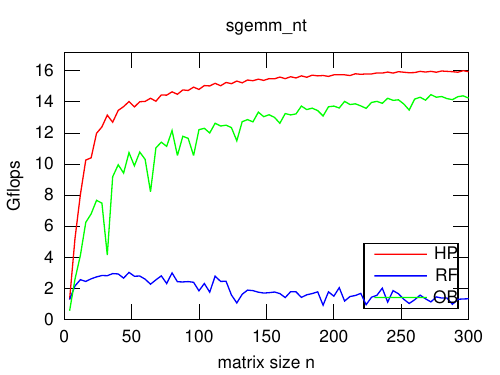} \label{fig:exp:cortex_a57:sgemm_nt}} 
\subfloat{\includegraphics[width=0.42\linewidth]{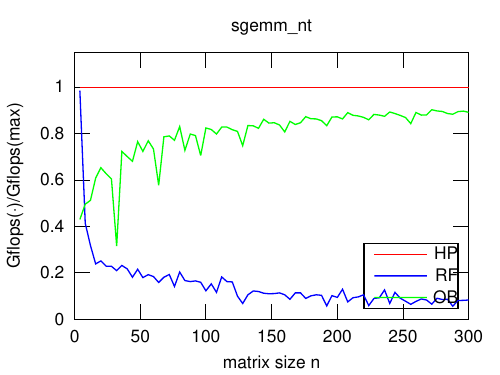} \label{fig:exp:cortex_a57:sgemm_nt_norm}} \\
\subfloat{\includegraphics[width=0.42\linewidth]{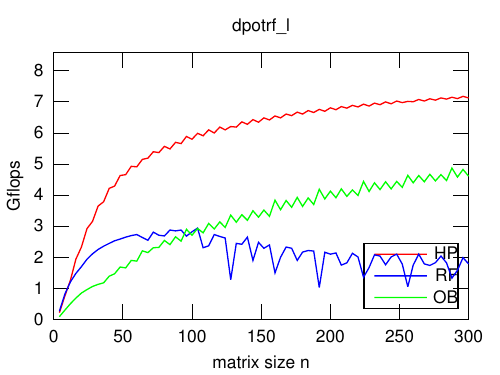} \label{fig:exp:cortex_a57:dpotrf}} 
\subfloat{\includegraphics[width=0.42\linewidth]{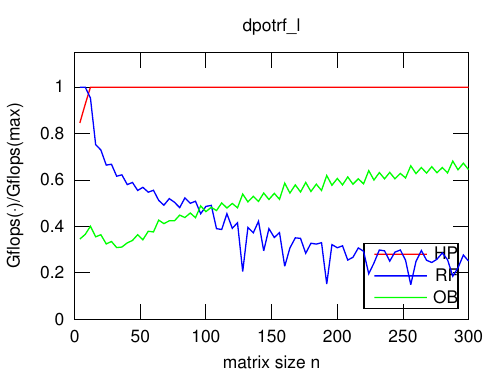} \label{fig:exp:cortex_a57:dpotrf_norm}} \\
\subfloat{\includegraphics[width=0.42\linewidth]{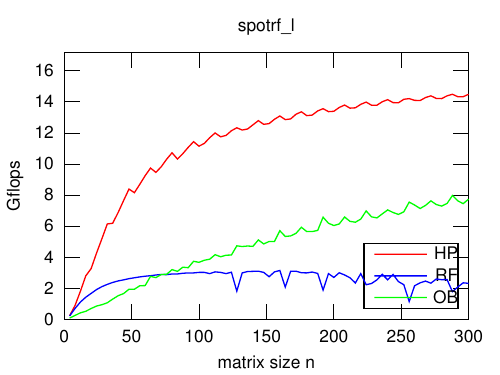} \label{fig:exp:cortex_a57:spotrf}} 
\subfloat{\includegraphics[width=0.42\linewidth]{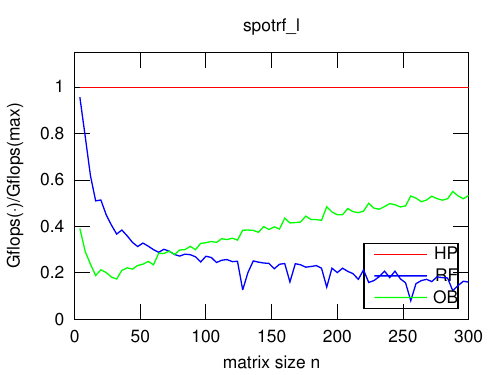} \label{fig:exp:cortex_a57:spotrf_norm}} \\
\caption{Performance of {\tt gemm\_nt} on ARM Cortex A57.}
\label{}
\end{figure}


\subsubsection{ARM Cortex A15} \label{sec:exp:blas:cortex_a15}

The ARM Cortex A15 is a relatively low-power architecture.
It is a 3-way superscalar architecture with out-of-order execution, but with a much smaller reorder buffer than Intel Haswell.
ARMv7A does not support vectorization in double precision: therefore, the scalar VFP instruction set is employed.
In single precision, it is possible to choose between the scalar VFP instruction set, or the 4-wide SIMD NEON instruction set.
Due to its high performance, BLASFEO HP employs the latter.
The Cortex A15 core can perform a 64-bit wide (double precision) and a 128-bit wide (single precision) FP multiplication-accumulate at every clock cycle, giving a throughput of 2 and 8 flops per cycle in double and single precision respectively.
{\color{black}
Each core has 32 KB 2-way associative instruction and data L1 caches.
All cores share a 16-way associative unified L2 cache.
The cache line size is 64 bytes.
}

In the implementation of BLASFEO HP, the panel size $p_s$ is 4 in both double and single precision.
The optimal BLASFEO HP {\tt gemm} kernel size is $4\times 4$ in double precision and $12\times 4$ in single precision.
Software prefetch has to be employed for both the left and the right factors, as there appear to be no hardware prefetch.

The test processor is the NVIDIA Tegra TK1 SoC (running at 2.3 GHz during all tests), equipped with 2 MB L2 cache.
The memory interface is 64-bit wide LPDDR3 giving 17 GB/s of bandwidth.
The amount of memory is 2 GB.

In the case of the ARM Cortex A15, only the {\tt gemm\_nt}
{\color{black}
and {\tt potrf\_l} routines are 
}
tested.

\paragraph{gemm\_nt}
In double precision, both BLASFEO HP and OpenBLAS perform well, very close to the maximum throughput.
BLASFEO RF and Eigen clearly suffer from the lack of software prefetch.
In single precision, the performance of BLASFEO HP clearly stands out.
OpenBLAS and BLASFEO RF do not employ vectorization, therefore losing a factor 4 with respect to BLASFEO HP.
Eigen appears to use vectorization, but its performance is about 2.5 times lower than BLASFEO HP.
{\color{black}
In both single and double precision, the small L1 cache associativity heavily penalizes BLASFEO RF for matrix sizes multiple of 32 (double precision) and 64 (single precision), due to the use of non-contiguous memory.
It is worthwhile to note that the panel-major format employed in BLASFEO HP gives very good performance also in the single precision case, where the $12\times 4$ {\tt gemm} kernel streams 4 panels at a time (3 L1-resident panels form $A$ and 1 L2-resident panel from $B$), despite the L1 cache being only 2-way associative.
This is a good example of the fact that the BLASFEO panel-major matrix format works very well in practice, despite being suboptimal due to the constraint of packing both $A$ and $B$ with the same panel size $p_s$ (equal to 4 in this case, while the optimal one would be equal to 12 for $A$ and 4 for $B$).
}

{\color{black}
\paragraph{potrf\_l}
The results are similar in the case of the Cholesky factorization {\tt potrf\_l}.
In double precision, only scalar instructions are used.
Therefore, compared to e.g. OpenBLAS, BLASFEO HP gives a slightly smaller speedup than in case of architectures with vector instructions, where the BLASFEO HP vectorized kernels have a large advantage over the scalar unblocked LAPACK routines.
In single precision, the speedup of BLASFEO HP is about 4 to 5 times, mainly due to the lack of vectorization in OpenBLAS.
}

\begin{figure}
\centering
\subfloat{\includegraphics[width=0.42\linewidth]{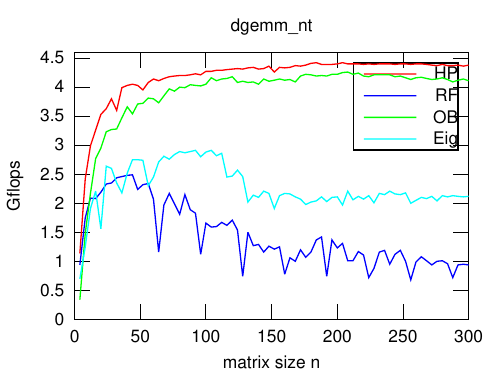} \label{fig:exp:cortex_a15:dgemm_nt}} 
\subfloat{\includegraphics[width=0.42\linewidth]{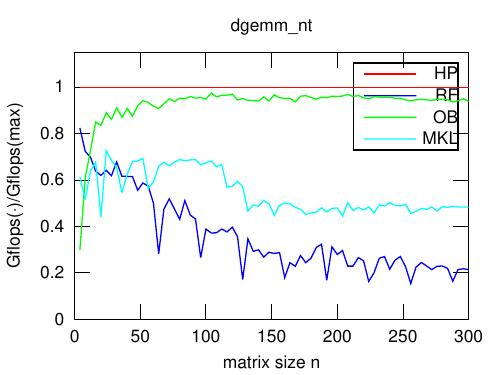} \label{fig:exp:cortex_a15:dgemm_nt_norm}} \\
\subfloat{\includegraphics[width=0.42\linewidth]{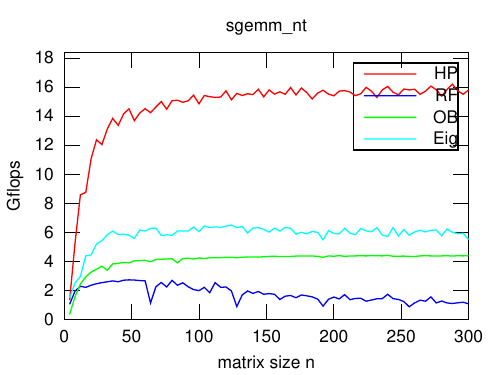} \label{fig:exp:cortex_a15:sgemm_nt}} 
\subfloat{\includegraphics[width=0.42\linewidth]{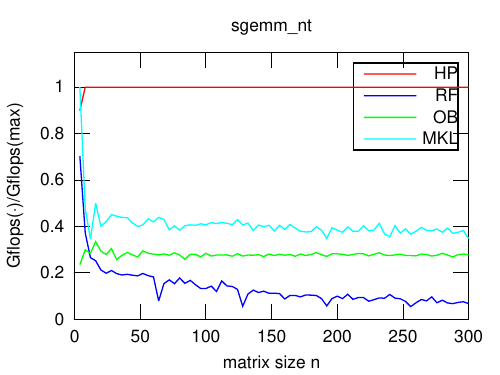} \label{fig:exp:cortex_a15:sgemm_nt_norm}} \\
\subfloat{\includegraphics[width=0.42\linewidth]{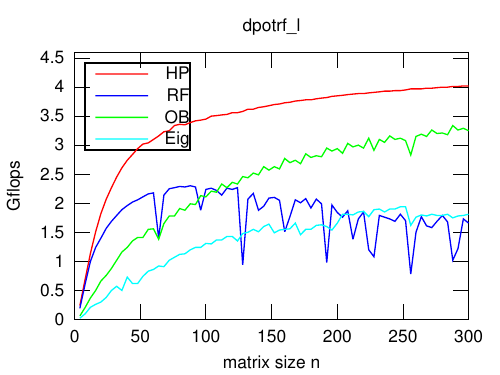} \label{fig:exp:cortex_a15:dpotrf}} 
\subfloat{\includegraphics[width=0.42\linewidth]{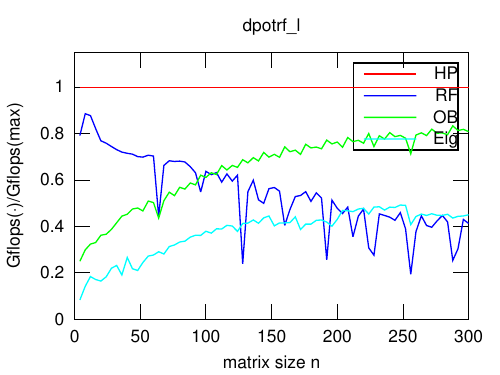} \label{fig:exp:cortex_a15:dpotrf_norm}} \\
\subfloat{\includegraphics[width=0.42\linewidth]{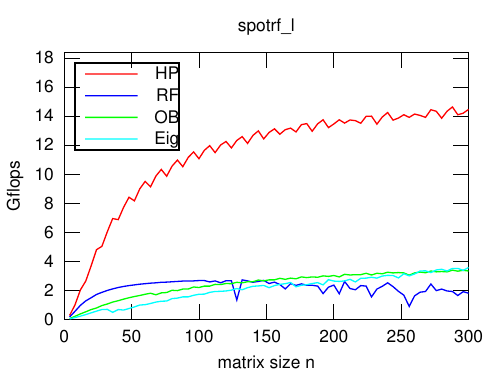} \label{fig:exp:cortex_a15:spotrf}} 
\subfloat{\includegraphics[width=0.42\linewidth]{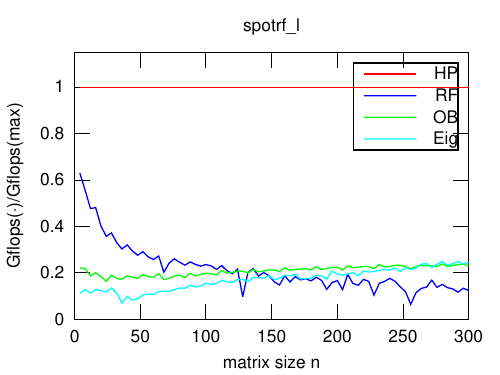} \label{fig:exp:cortex_a15:spotrf_norm}} \\
\caption{Performance of {\tt gemm\_nt} on ARM Cortex A15.}
\label{}
\end{figure}


\subsection{Backward Riccati recursion} \label{sec:exp:hpmpc}

The backward Riccati recursion is a special structured factorization for the KKT matrix arising in optimal control problems.
The recursion reads
\begin{equation*}
P_n = Q + A P_{n+1} A^T - (S + A P_{n+1} B^T) (R + B P_{n+1} B^T )^{-1} (S^T + B P_{n+1} A^T)
\end{equation*}
where the matrices $P_{n+1}$ and $\begin{bmatrix} R & S \\ S^T & Q \end{bmatrix}$ are assumed to be symmetric positive definite.
The matrices $A$, $Q$, $P$ have size $n_x\times n_x$, the matrices $B$ and $S^T$ have size $n_x\times n_u$ and the matrix $R$ has size $n_u\times n_u$, where $n_x$ is the number of states and $n_u$ is the number of controls of the system.
The recursion is repeated $N$ times, where $N$ is the control horizon length.

The Riccati recursion can be implemented efficiently as~\cite{Frison2014}
\begin{align*}
C \gets & \begin{bmatrix} B^T \\ A^T \end{bmatrix} \cdot \mathcal L_{n+1} \\
\begin{bmatrix} \Lambda_n & 0 \\ L_n & \mathcal L_n \end{bmatrix} \gets & \left( C \cdot C^T \right) ^{\tfrac 1 2}
\end{align*}
where $\mathcal L_{n+1}$ is the lower Cholesky factor of $P_{n+1}$ and the exponent $^{\tfrac 1 2}$ denotes the lower triangular Cholesky factorization.
The algorithm can be implemented using the {\tt trmm\_rlnn} and {\tt syrk\_ln} BLAS routines and the {\tt potrf\_l} LAPACK routine.
Note that this algorithm gives the opportunity to fuse the {\tt syrk\_ln} and the {\tt potrf\_l} routines.

\begin{figure}
\centering
\subfloat{\includegraphics[width=0.42\linewidth]{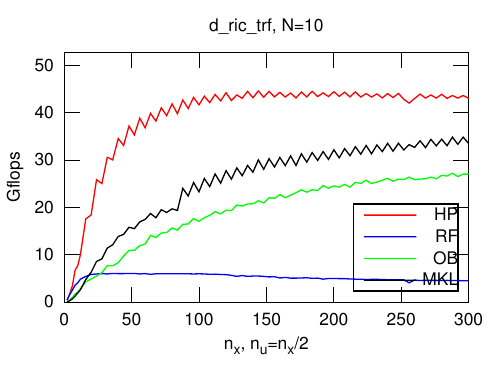} \label{fig:exp:haswell:d_ric_trf}} 
\subfloat{\includegraphics[width=0.42\linewidth]{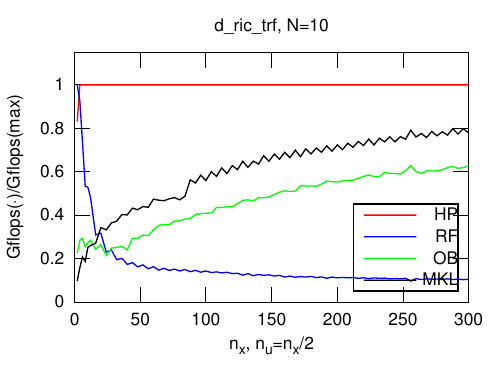} \label{fig:exp:haswell:d_ric_trf_norm}} \\
\caption{Performance of {\tt d\_ric\_trf} on Intel Haswell.}
\label{fig:ric}
\end{figure}

The computational performance of the algorithm is shown in Figure~\ref{fig:ric}, which closely resembles the performance plots of BLAS and LAPACK routines.
Therefore, also in this case the BLASFEO HP is the best choice for the matrix sizes of interest, giving a speed-up of about 2-3 times for a number of states $n_x$ up to 100, with respect to optimized BLAS and LAPACK implementations.

\subsection{Dual Newton strategy} \label{sec:exp:qpdunes}

Aim of this section is to demonstrate how the linear algebra provided in BLASFEO can enhance the performance of new or existing embedded optimization tools. 
As an example, we take the open-source software qpDUNES~\cite{Frasch2015}, a dual Newton strategy for QPs arising in optimal control. 

The main idea of the algorithm is to introduce Lagrange multipliers for the equality constraints imposed by the system dynamics and solve the resulting (unconstrained) dual problem with Newton's method. 
In this framework, one of the most computationally expensive operations per iteration of the solver is the solution of a linear system, requiring the factorization of the dual Hessian $H_d$ and the computation of the Newton direction. 
The matrix $-H_d$ is positive definite and has a block tri-diagonal structure with $N$ diagonal blocks. 
This motivates the use of a block banded Cholesky factorization, which has a complexity that scales linearly in the number of blocks.

\begin{figure}
	\centering
	\includegraphics[width=0.84\linewidth]{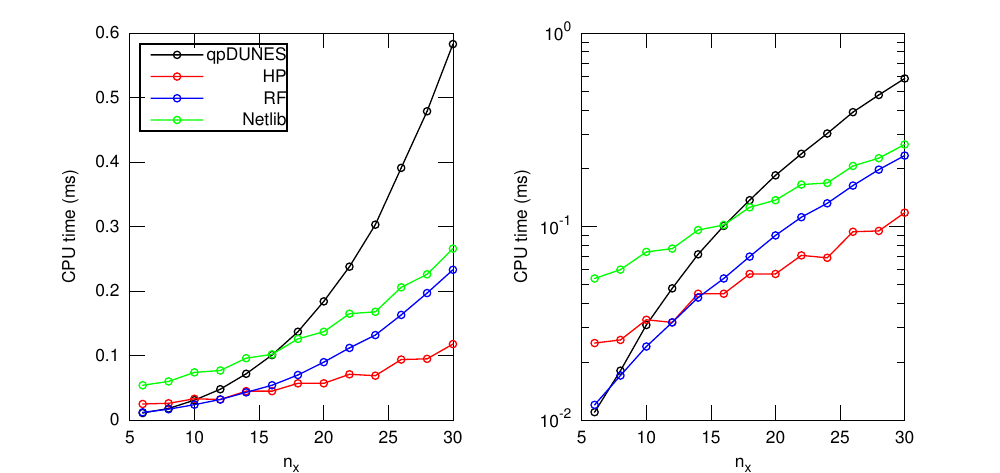}
	\caption{CPU timings for the computation of Newton direction, varying $n_x$. Comparison of the existing qpDUNES implementation against the re-implementation using BLASFEO (HP, RF and WR linked to Netlib BLAS and LAPACK).}
	\label{fig:qpdunes1}
\end{figure}

To show the room for improvement on the software by the use of BLASFEO, we replace all operations in the factorization and substitution steps of qpDUNES with calls to BLASFEO subroutines. 
{\color{black}
The benchmark problem is the control of a linear chain of masses and springs.
The number of masses (and therefore the number of states $n_x$ and controls $n_u$) as well as the control horizon length $N$ can be freely scaled.
The Hessian matrices are diagonal, and only bounds on the state and control variables are considered (and not general polytopic constraints).
In this test,
}
the horizon length and number of controls are kept constant with values $N = 20$ and $n_u = 2$ respectively,
{\color{black}
while the number of states $n_x$ is varied between 6 and 30.
}
Note that $n_u$ does not affect the timings since all Hessian blocks are of size $n_x \times n_x$. 
The results are shown in Figure~\ref{fig:qpdunes1}. 
BLASFEO RF matches the performance of the existing implementation for the smaller sizes, while BLASFEO HP is over 5 times faster for the largest sizes.


Note that the time needed to convert the data between the row major matrix format (qpDUNES internal format) and the BLASFEO structure formats is included in the timings of the BLASFEO implementations.
Since only the solution of the Newton system is optimized, the overall speedup of the software is lower due to Amdahl's law. 
However, the results indicate that using BLASFEO throughout the code can lead to significant performance gains.

\section{Conclusion} \label{sec:con}

This paper presented the implementation details of BLASFEO, a library of BLAS- and LAPACK-like routines optimized for use in embedded optimization.
As a key difference with respect to highly-tuned BLAS and LAPACK routines, BLASFEO is designed to give the best performance for rather small matrices that fit in some level of cache.
Compared to the best open-source and proprietary single-threaded BLAS and LAPACK libraries, the HP implementation of BLASFEO shows large speedups for all the matrix sizes tested in this paper, i.e. for sizes up to 300.
Therefore BLASFEO shows that it is possible to employ high-performance techniques for the implementation of DLA routines optimized for small matrix sizes.

For matrices of size up to 100, BLASFEO HP shows a speedup of about 20-30\% in the case of level 3 BLAS-like routines, and of about 2-3 times in the case of LAPACK-like routines, compared to the best available DLA implementations.
In case of BLAS-like routines, the speedup is mainly due to the fact that the panel-major matrix format is exposed to the user, and therefore to the fact that on-line packing of matrices is avoided.
In case of LAPACK-like routines, the much larger speed-up is mainly due to their implementation as if they were BLAS-like routines. 
The BLAS routines and the unblocked LAPACK routines in the standard LAPACK implementation are both replaced with tailored DLA kernels implemented using register blocking and vectorization.
This greatly enhances performance for small matrices, and it could be used as a technique to implement higher-performing standard LAPACK routines.

{\color{black}
Currently the high-performance version of BLASFEO only provides single-threaded DLA routines.
Future work will focus on multi-threaded routines and investigate the parallelization level achievable in the BLASFEO framework.
}


\appendix

\section{BLASFEO API} \label{sec:inter}

BLASFEO has a different API than BLAS and LAPACK.
This is due to the need of dealing with different matrix formats (column-major for BLASFEO RF and BLASFEO WR, and panel-major for BLASFEO HP).
In C, this can be achieved by means of structures.
BLASFEO provides macros to easily access matrix and vector elements without having to know their storage format.

\subsection{Matrix and vector structures} \label{sec:inter:str}

BLASFEO employs C structures to handle matrices and vectors.
Some structure members are common to all BLASFEO implementations (and would be public members in a C++ object), others are used specifically in some implementations (and would be private members in a C++ object).
C structures do not easily allow the use of private members, so all members are 'public'.
In the following only common members are described.

\subsubsection{Definition of structures} \label{sec:inter:str:def}

The structure defining a double precision matrix is {\tt blasfeo\_dmat} and it is defined as 
\begin{verbatim}
struct blasfeo_dmat{
    double *pA;
    int m;
    int n;
    int memsize;
}
\end{verbatim}
where: 
\begin{description}
\item[pA] is a pointer to the first element of the matrix
\item[m] is the number of rows of the matrix
\item[n] is the number of columns of the matrix
\item[memsize] is the size in bytes of the chunk of memory addressed by the structure.
\end{description}
The structure defining a single precision matrix is {\tt blasfeo\_smat} and it is defined similarly.

The structure defining a double precision vector is {\tt blasfeo\_dvec} and it is defined as
\begin{verbatim}
struct blasfeo_dvec {
    double *pa;
    int m;
    int memsize;
}
\end{verbatim}
where 
\begin{description}
\item[pa] is a pointer to the first element of the vector
\item[m] is the number of elements of the vector
\item[memsize] is the size in bytes of the chunk of memory addressed by the structure.
\end{description}
The structure defining a single precision vector is {\tt blasfeo\_svec} and it is defined similarly.

Vectors are contiguous pieces of memory: in the language of standard BLAS, they have unit stride.
This prevents level 2 routines in BLASFEO to be employed to operate on e.g. rows and columns of matrices.
In order to do so, a set of routines to extract and inserts rows and columns of matrices as vectors is provided.

\subsubsection{Memory usage} \label{sec:inter:str:man}

The amount of memory needed to store a matrix in panel-major format is generally not the same as in the column-major format.
Since matrix and vector structures hide all implementation details, the equivalents of C++ constructor/destructor functions are provided.
In order to guarantee embeddability and avoid system calls on the critical path, the option to avoid dynamic memory allocation is provided.
In BLASFEO, all matrix and vector structures can be created using either dynamic memory allocation or externally provided memory.
The former method provides ease of use and it is ideal for prototyping or debugging. 
The latter method avoids any internal memory allocation and employs memory externally allocated (either automatically, statically or dynamically).
Therefore it is ideal for performance (no system calls are performed) and embeddability (any type of memory allocation supported by the system can be employed).
As a drawback, the user has to take care of possibly present alignment requirements.

\paragraph{Dynamically allocated memory} The routine
\begin{verbatim}
void blasfeo_allocate_dmat(int m, int n, struct blasfeo_dmat *sA);
\end{verbatim}
where
\begin{description}
\item[m] is the number of rows of the matrix
\item[n] is the number of columns of the matrix
\item[sA] is a pointer to the structure {\tt blasfeo\_dmat} to be created,
\end{description}
sets the members of a {\tt blasfeo\_dmat} structure and dynamically allocates {\tt sA.memsize} bytes of memory for internal use (automatically taking care of alignment requirements).
The routine
\begin{verbatim}
void blasfeo_free_dmat(struct blasfeo_dmat *sA);
\end{verbatim}
where
\begin{description}
\item[sA] is a pointer to the structure {\tt blasfeo\_dmat} to be destructed
\end{description}
takes care of freeing the memory dynamically allocated in a previous call to {\tt blasfeo\_allocate\_dmat}.
Analogue routines exist for single precision and vectors.

\paragraph{Externally provided memory} The routine
\begin{verbatim}
int blasfeo_memsize_dmat(int m, int n);
\end{verbatim}
where
\begin{description}
\item[m] is the number of rows of the matrix
\item[n] is the number of columns of the matrix
\end{description}
returns the number of bytes of memory that have to be provided for a {\tt blasfeo\_dmat} structure holding a matrix of size $m\times n$.
The routine
\begin{verbatim}
void blasfeo_create_dmat(int m, int n, struct blasfeo_dmat *sA, void *memory);
\end{verbatim}
where
\begin{description}
\item[m] is the number of rows of the matrix
\item[n] is the number of columns of the matrix
\item[sA] is a pointer to the structure {\tt blasfeo\_dmat} to be created
\item[memory] is a pointer to a properly aligned chunk of at least {\tt blasfeo\_memsize\_dmat(m, n)} bytes of memory,
\end{description}
sets the members of a {\tt blasfeo\_dmat} structure using the externally allocated memory passed through the argument {\tt memory}.
The argument {\tt memory} is unchanged on exit.
Since there is no internal memory allocation, there is no analogous to the {\tt blasfeo\_free\_dmat} routine.

The memory addressed by {\tt memory} is not touched by the routine, meaning that the routine can be used to cast a piece of memory as a {\tt blasfeo\_dmat} structure (provided that the user is aware of the internal matrix layout).

If a large chunk of memory is allocated to be used with several matrices at once, the member {\tt memsize} returns the minimum number of bytes that have to be reserved for internal use in each {\tt blasfeo\_dmat} structure.
The value of {\tt memsize} is a multiple of the minimum alignment required for use in the {\tt blasfeo\_dmat} structure.
Therefore, if the pointer {\tt memory} is properly aligned, it is still properly aligned if moved of {\tt memsize} bytes.

Analogue routines exist for vectors (wit name ending with {\tt \_dvec}) and single precision ({\tt smat} and {\tt svec}).

\subsection{Interface of linear algebra routines} \label{sec:inter:api}

BLASFEO comes with its own API to linear algebra routines.
The API somehow resembles the standard BLAS and LAPACK API, but with some important differences.
A generic linear algebra routine in BLASFEO looks like
\begin{verbatim}
return_value = routine_name( operation_size_1, ..., operation_size_m, \
                             operand_1, ..., operand_n );
\end{verbatim}
where the first arguments are of type {\tt int} and define the operation size, and the last arguments define the operands of the operation.
Operands can be
\begin{description}
\item[scalar:] a scalar operand is either a {\tt double} or a {\tt float}.
\item[vector:] a vector operand is in the form \{{\tt blasfeo\_dvec *sx, int xi}\} in double precision, and similarly for single precision.
The integer {\tt xi} defines the position of the first element of the sub-vector that is the actual operand.
\item[matrix:] a matrix operand is in the form \{{\tt blasfeo\_dmat *sA, int ai, int aj}\} in double precision, and similarly for single precision.
The integers {\tt ai} and {\tt aj} define the position (given as row and column indices respectively) of the first element of the sub-matrix that is the actual operand.
\end{description}

BLAS and LAPACK {\tt char} arguments (describing options for e.g. upper/lower, nontransposed/transposed, right/left, \dots) are hard-coded in the {\tt routine\_name}, slightly reducing routine overhead.

Linear algebra routines in BLASFEO are non-destructive, meaning that there is an argument reserved for the output operand (even if, in some cases, the same structure can be used for input and output operands).
In many cases this avoids the need to explicitly perform a matrix or vector copy, helping to reduce overhead for small matrices.

\subsubsection{Example} \label{sec:inter:api:ex}

\paragraph{gemm} The interface for the double-precision general matrix-matrix multiplication (nontransposed-transposed version) is
\begin{verbatim}
void blasfeo_dgemm_nt(int m, int n, int k, \
    double alpha, \
    struct blasfeo_dmat *sA, int ai, int aj, \
    struct blasfeo_dmat *sB, int bi, int bj, \
    double beta, \
    struct blasfeo_dmat *sC, int ci, int cj, \
    struct blasfeo_dmat *sD, int di, int dj);
\end{verbatim}
implementing the operation
\begin{equation*}
D \gets \alpha \cdot A \cdot B^T + \beta \cdot C
\end{equation*}
where $A$ is a matrix of size $m\times k$, $B$ is a matrix of size $n\times k$, and $C$ and $D$ are matrices of size $m\times n$.
The matrices $C$ and $D$ can coincide.

\subsection{List of currently available routines} \label{sec:inter:rout}

{\color{black}
BLASFEO is a young and dynamic project, and the list of implemented routines is constantly increasing.
The priority in the implementation of new routines follows the needs for features of other software packages, as well as user requests.

The process of adding a new routine starts with the BLASFEO WR and RF implementations, followed by a generic C version of the DLA kernels in BLASFEO HP.
The implementation of assembly coded DLA kernels for the supported architectures follows, again trying to meet the needs of applications and users.

At the time of writing this paper, the following DLA routines are available in BLASFEO, for real floating point numbers in both single and double precision (complex data types are not targeted in BLASFEO, since they are of limited interest in embedded optimization applications):
\begin{description}
\item[LAPACK:] {\tt potrf\_l, getrf\_\{pivot, nopivot\}, geqrf, gelqf, syrk\_potrf\_ln}
\item[BLAS level 3:] {\tt gemm\_\{dn, nd, nn, nt\}, syrk\_ln, trmm\_\{rlnn, rutn\}, trsm\_\{llnu, lunn, rltn, rltu, rutn\} }
\item[BLAS level 2:] {\tt gemv\_\{d, n, nt, t\}, symv\_l, trmv\_\{lnn, ltn, unn, utn\}, trsv\{lnn, lnu, ltn, ltu, unn, utn\} }
\item[BLAS level 1:] {\tt axpy, axpby, dot, rotg, \{col, row\}rot }
\end{description}
Note that some variants are not covered in standard BLAS and LAPACK (e.g. {\tt d} indicates for diagonal matrices).

Additionally, BLASFEO provides a rich set of auxiliary routines, which cover operations not traditionally included in BLAS and LAPACK.
These routines comprise:
\begin{itemize}
\item creation/allocation/deallocation/packing of matrices/vectors
\item copy/transpose/scale/sum/permutation of generic/triangular matrices
\item insertion/extraction/scale/sum/permutation of rows/columns/diagonals/elements of matrices
\end{itemize}

Overall, all these routines allow to implement a large set of algorithms using BLASFEO as an efficient and convenient framework.
}

\section{Netlib-style linear algebra and compilers} \label{sec:compilers}

{
\color{black}

\begin{figure}
\centering
\subfloat[]{\includegraphics[width=0.42\linewidth]{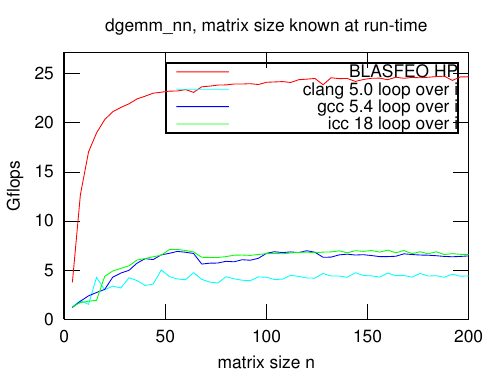}
\label{fig:exp:sandybridge:codegen:dgemm_dyn}}
\subfloat[]{\includegraphics[width=0.42\linewidth]{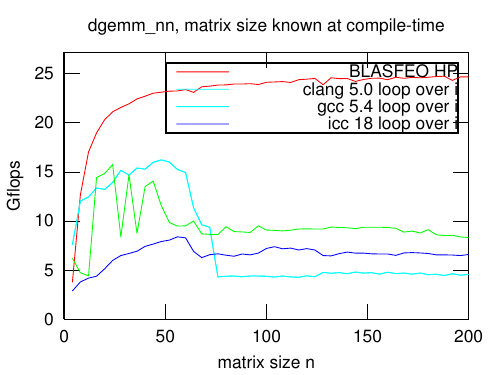}
\label{fig:exp:sandybridge:codegen:dgemm}}
\\
\subfloat[]{\includegraphics[width=0.42\linewidth]{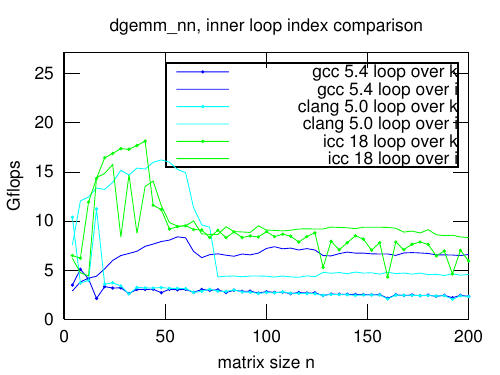}
\label{fig:exp:sandybridge:codegen:dgemm_loop_ik}}
\subfloat[]{\includegraphics[width=0.42\linewidth]{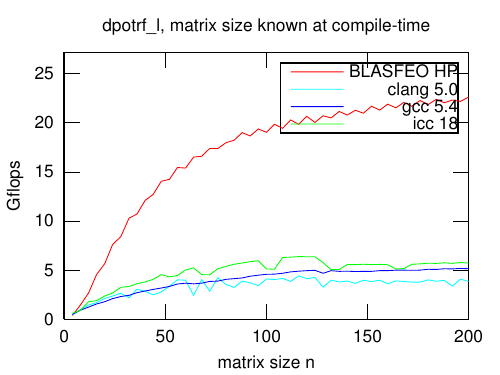} 
\label{fig:exp:sandybridge:codegen:dpotrf}}
\caption{Performance of compiled code for {\tt gemm\_nn} and {\tt potrf\_l} on Intel Ivy-Bridge.}
\label{fig:exp:sandybridge:codegen}

\end{figure}
\begin{figure}
\centering
\subfloat[]{\includegraphics[width=0.42\linewidth]{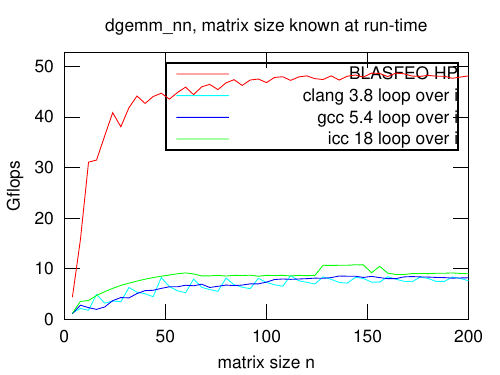}
\label{fig:exp:haswell:codegen:dgemm_dyn}}
\subfloat[]{\includegraphics[width=0.42\linewidth]{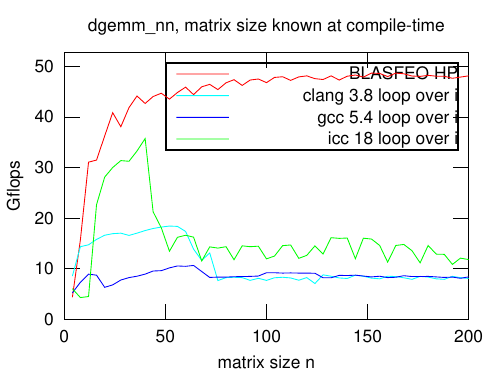}
\label{fig:exp:haswell:codegen:dgemm}}
\\
\subfloat[]{\includegraphics[width=0.42\linewidth]{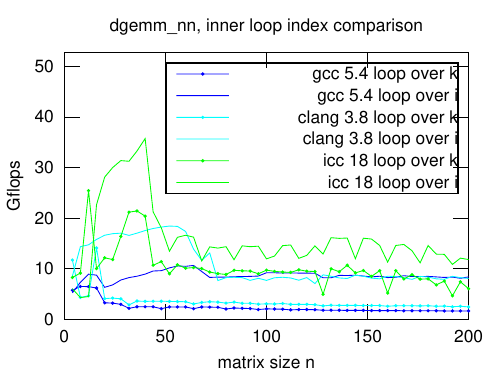}
\label{fig:exp:haswell:codegen:dgemm_loop_ik}}
\subfloat[]{\includegraphics[width=0.42\linewidth]{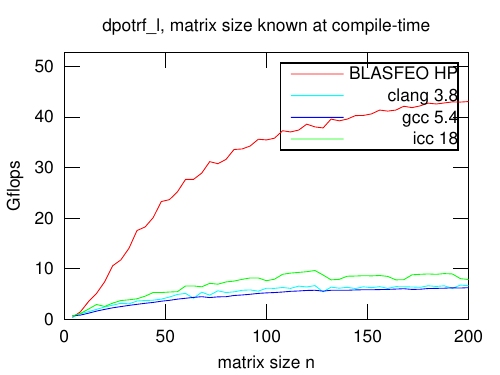} 
\label{fig:exp:haswell:codegen:dpotrf}}
\caption{Performance of compiled code for {\tt gemm\_nn} and {\tt potrf\_l} on Intel Haswell.}
\label{fig:exp:haswell:codegen}
\end{figure}

This section presents a performance comparison between BLASFEO HP and a plain C implementation of two BLAS routines, compiled by some of the most common optimizing compilers ({\tt gcc}, {\tt clang} and {\tt icc}).

\subsection{Routines Implementation}

The tested routines are: 
\begin{itemize}
\item \texttt{dgemm\_nn} This is the general matrix-matrix multiplication, double precision version (where the factor matrices A and B are non transpose, non transpose respectively).
This is the most common level 3 BLAS routine and the most straightforward to implement.
\item \texttt{dpotrf\_l} This is the Cholesky decomposition, double precision (where the lower triangular factor is computed).
This routine is chosen in order to check compiler behavior on less-trivial cases (triangular matrix, dependency between loop boundaries).
\end{itemize}

For the purpose of this test, these routines have been coded as triple-loop LA routines in reference Netlib style using plain C.
All matrices are store in column-major.


The \texttt{gemm} routine has been implemented in two different variants, with innermost loop over \texttt{k} or over \texttt{i}.
The former is the variant used in hand-optimized {\tt gemm} kernels, while the latter variant should be easier to auto-vectorize by simply unrolling the innermost loop.

Furthermore all variants have been compiled in two different ways, one in which matrix sizes are known at compile-time, and one in which they are only known at run-time.

\subsection{Platform description}

The compilers performance is tested on two different architecture: Intel Sandy-Bridge and Intel Haswell.

The former architecture is implemented in the processor Intel Core i3 3420 (Ivy-Bridge), running at 3.4 GHz during all tests.
The architecture supports the AVX ISA.
The maximum throughput in double precision is 27.2 Gflops.

The latter architecture is implemented in the processor Intel Core i7 4800MQ (Haswell), running at 3.3 GHz during all tests.
The architecture supports the AVX2 and FMA ISAs.
The maximum throughput in double precision is 58.2 Gflops.

\subsection{Compilers and compilation flags}

Different compilers with the same set of compilation flags have been tested.

The compilers included in this comparison are:

\begin{itemize}
\item \texttt{GCC 5.4}
\item \texttt{Clang/LLVM 3.8}
\item \texttt{Intel C Compiler 18 (2018.0.128)}
\end{itemize}

The following compilation flags have been used:

\begin{itemize}
\item \texttt{-O3} Sets the optimization level.
\item \texttt{-funroll-loops} Tells the compiler to unroll all the loops when profitable. Loop unrolling avoids loop related overhead (loop index handling and branch misprediction).
\item \texttt{-m64} Generates code for x86-64 architecture.
\item \texttt{-mavx} ({\tt gcc} and {\tt clang} only) Allows the compiler to use AVX ISA (Sandy-Bridge architecture).
\item \texttt{-mavx2 -mfma} ({\tt gcc} and {\tt clang} only) Allows the compiler to use AVX2 and FMA ISAs (Haswell architecture).
\item \texttt{-fno-alias} ({\tt icc} only) The Assume No Aliasing option tells the compiler that the data matrices are not aliased.
\item \texttt{-fno-fnalias} ({\tt icc} only) The Assume Aliasing Across Function Calls option tells the compiler that no aliasing occurs within function bodies but might occur across function calls.
\item \texttt{-fast} ({\tt icc} only) Forces static link, allows interprocedural optimization and more aggressive floating point optimizations. Generates instructions for the highest instruction set on the compilation host processor.
\end{itemize}

\subsection{Numerical Experiments}

Figures \ref{fig:exp:haswell:codegen:dgemm_dyn} and \ref{fig:exp:sandybridge:codegen:dgemm_dyn} show the performance of the \texttt{dgemm} variant with inner loop over {\tt i}, when matrices sizes, thus loops bounds, are known only at run time.
Performance is rather poor for all compilers, floating around 20\% of maximum throughput.

Figures \ref{fig:exp:haswell:codegen:dgemm} and \ref{fig:exp:sandybridge:codegen:dgemm} show the performance of the \texttt{dgemm} variant with inner loop over {\tt i}, when loops bounds are already known at compile time.
In this case \texttt{icc} and \texttt{clang} achieve better results thanks to loop unrolling and vectorization.
Their performance is close to the one achieved by BLASFEO HP for matrix sizes of a few units.
However, as soon as the matrix sizes reach a few tens, BLASFEO HP performance is noticeable better.
The performance gap gets very large once the memory footprint exceeds L1 cache size, as in that case the performance of all compiler versions drops considerably.
This sharp drop in performance 
hints that the compilers do not introduce cache blocking.
It has to be noted that \texttt{clang} needs to be aware that matrix pointers refer to disjointed memory locations in order to properly perform loop unrolling and vectorization.
This information is passed to the compiler with the C type qualifier keyword \texttt{restrict}.
Also \texttt{icc} receives this information through the compilation flags \texttt{no-alias} and \texttt{no-fnalias}.
It has been observed that for \texttt{icc} (conversely to \texttt{clang}) that information have little impact on performance, as the compiler is still able to vectorize  even if the flag is not present.
The {\tt gcc} compiler does not benefit much from this additional information.

Figures \ref{fig:exp:haswell:codegen:dgemm_loop_ik} and \ref{fig:exp:sandybridge:codegen:dgemm_loop_ik} show the
comparison of the performance of {\tt dgemm} variants with inner loop over \texttt{i} and over \texttt{k}, for both matrix sizes known at run and at compile time.
On average compilers perform better if the inner loop is over \texttt{i}. This is expected since this case can be easily  vectorized by unrolling only the innermost loop.
However this variant has a lower arithmetic intensity than the variant with inner loop over \texttt{k}.
Namely, at each iteration of the innermost loop, the version with inner loop over \texttt{i} requires 3 memops (2 \texttt{load} and 1 \texttt{store}) every 2 flops (1 \texttt{mul} and 1 \texttt{add}), while the version with inner loop over {\tt i} only requires 2 memops (2 loads) every 2 flops.

Since the version with inner loop over {\tt k} has higher arithmetic intensity, it is generally preferred in optimized BLAS implementations.
However, it appears to be more difficult to autovectorize for the compiler, since this requires the two outermost loops to be partially unrolled.
The \texttt{icc} compiler is an exception, as it performs in a similar way on both variants. 

Figures \ref{fig:exp:haswell:codegen:dpotrf} and \ref{fig:exp:sandybridge:codegen:dpotrf}
show the Cholesky decomposition, chosen as an example of a more complex routine that is more more difficult to optimize.
These numerical experiments show that compilers give poor performance also if matrix sizes are know at compile time.

In some figures performances drops are clearly visible when the matrix size is multiple of 16 or 32. This effect is due to the limited associativity of caches, when memory is not accessed in a continuous way (as e.g. the case if elements of a column-major matrix are accessed by rows).

}





\end{document}